\documentclass[twocolumn,superscriptaddress,amsfont,amssymb,amsmath, showpacs,balancelastpage, nofootinbib]{revtex4-1}
\usepackage{graphicx,longtable,mathrsfs,color,array}
\usepackage[hidelinks]{hyperref}
\usepackage[usenames,dvipsnames]{xcolor}
\usepackage{amssymb,amsmath,mathtools,mathrsfs}
\usepackage{epsfig,subfigure,placeins,float}
\usepackage{booktabs,longtable,ctable,multirow}
\usepackage{exscale,relsize}
\usepackage[normalem]{ulem}
\usepackage{enumerate}
\usepackage{times, mathptmx}
\newcommand{\nv}{\hat{\bf n}}
\newcommand{\jcap}{JCAP}
\newcommand{\mnras}{MNRAS}
\newcommand{\aap}{A\&A}

\newcommand{\apjs}{ApJS}

\newcommand{\procspie}{Proceedings of the SPIE}
\newcommand{\physrep}{Physics Reports}
\newcommand{\hiclass}{{\tt hi\_class}}
\newcommand{\uKam}{\mu\text{K-arcmin}}
\newcolumntype{C}[1]{>{\centering\let\newline\\\arraybackslash\hspace{0pt}}m{#1}}

\begin{document}
\title{The Observational Future of Cosmological Scalar-Tensor Theories}
\author{D.~Alonso}
\affiliation{University of Oxford, Denys Wilkinson Building,
  Keble Road, Oxford, OX1 3RH,  UK}
  \author{ E.~Bellini}
\affiliation{University of Oxford, Denys Wilkinson Building,
  Keble Road, Oxford, OX1 3RH,  UK}
  \affiliation{ICCUB, University of Barcelona, (IEEC-UB), Mart\'i i Franqu\`es, E08028 Barcelona, Spain}
  \author{P.~G.~Ferreira}
  \affiliation{University of Oxford, Denys Wilkinson Building,
  Keble Road, Oxford, OX1 3RH,  UK}
  \author{M.~Zumalac\'arregui}
  \affiliation{Nordita, KTH Royal Institute of Technology and Stockholm University,
Roslagstullsbacken 23, SE-106 91 Stockholm, Sweden}
\affiliation{Berkeley Center for Cosmological Physics and University of California at Berkeley, CA94720, USA}

\begin{abstract}
  The next generation of surveys will greatly improve our knowledge of cosmological gravity. In this paper we focus on how Stage IV  photometric redshift surveys, including weak lensing and multiple tracers of the matter distribution and radio experiments combined with measurements of the cosmic microwave background will lead to precision constraints on deviations from General Relativity. We use a broad subclass of Horndeski scalar-tensor theories to forecast the accuracy with which we will be able to determine these deviations and their degeneracies with other cosmological parameters.  Our analysis includes relativistic effects, does not rely on the quasi-static evolution and makes conservative assumptions about the effect of screening on small scales. We define a figure of merit for cosmological tests of gravity and show how the combination of different types of surveys, probing different length scales and redshifts, can be used to pin down constraints on the gravitational physics to better than a few percent, roughly an order of magnitude better than present probes. Future cosmological experiments will be able to constrain the Brans-Dicke parameter at a level comparable to Solar System and astrophysical tests.
 \end{abstract}

  \date{\today}
  \maketitle

\section{Introduction}\label{sec:intro}
The future of observational cosmology holds great promise. Not only will it be possible to refine our understanding of many of the ingredients of our current standard cosmological model (such as, for example, the geometry of space-time or the constituents of the mass density of Universe), but these observations will also allow us to explore truly foundational aspects in physics. In particular, it should be possible to test some of the most essential assumptions that go into constructing our working relativistic model of the Universe. One of the most fundamental assumptions is that gravity is perfectly described by Einstein's General Theory of Relativity.

General Relativity (GR) has been exquisitely tested on non-cosmological scales in the weak-field limit (see \cite{Berti:2015itd} for a recent review) and, more recently, has been shown to be broadly consistent with the first direct detection of gravitational waves from the merger of two black holes \cite{Abbott:2016blz}. The common-sense approach has been to assume that we can safely extrapolate GR to cosmological scales and proceed to accurately calculate various properties of the Universe. This is, of course, an ambitious extrapolation, over fifteen orders of magnitude in length scale and probing a completely new regime of gravitational potential and curvature \cite{Baker:2014zba}. Furthermore, there are a number of alternative theories that are consistent with non-cosmological tests but which can lead to a variety of cosmic histories \cite{Clifton:2011jh}. Given all of this, it makes sense to construct cosmological tests of gravity which will be able to distinguish between GR and its alternatives.

The time is right to do so. A battery of large scale surveys are planned to roll out over the next decade. More specifically we expect a new generation of photometric and spectroscopic galaxy redshift 
surveys, weak-lensing surveys, continuum surveys of radio galaxies, intensity mapping surveys of 
neutral hydrogen (HI), and a concerted campaign to map the temperature and polarization of the cosmic microwave background. These surveys will access a wide range of redshifts and scales
with different sensitivities and will be affected by different systematic uncertainties. Cross-correlating these surveys will allow us to mitigate these effects and extract precise information on the morphology
and evolution of large scale structure. Indeed, as was shown in \cite{Yoo:2012se,Alonso:2015uua,Alonso:2015sfa,Fonseca:2015laa}, combinations of various future data sets can be used to great effect to detect horizon-scale, general relativistic corrections to cosmological observables.

The next generation of surveys will deliver further tests of fundamental physics beyond the nature of gravity. For example, an important program in this direction is the determination of the mass scale of neutrinos, which is, arguably, easier to read off from their imprint on large scale structure and the cosmic microwave background (CMB) \cite{Ade:2015xua} than from laboratory experiments \cite{Agashe:2014kda}. The minimal mass scale of neutrinos ($\sum m_{\nu,i}\gtrsim 0.06 eV$ in the normal hierarchy) is set by oscillation experiments. Despite its smallness, it falls in the range of sensitivity for future experiments \cite{Font-Ribera:2013rwa,Allison:2015qca} and it is hence necessary to include it as a cosmological parameter. This is even more important in tests of gravity, whose signatures have been shown to be degenerate with neutrino masses in a range of models \cite{Barreira:2014jha,Baldi:2013iza,Baldi:2016oce}.

In this paper we look forward and forecast how this step changes in observational cosmology will affect our understanding of cosmological gravity. In broad terms, we will try to understand how our constraints on GR will improve over the next decade and how they will compare with constraints on other length scales. Ideally we would like to be completely agnostic about the theory of gravity. That is, we would like to make as few assumptions as possible about the space of theories within which we explore deviations. There has been great progress in characterizing gravity on linearly perturbed cosmological scales, from completely general approaches \cite{Amin:2007wi,Pogosian:2010tj,Bean:2010zq,Dossett:2011tn,Kunz:2006ca,Baker:2011jy,Baker:2012zs,Battye:2012eu,Battye:2013ida} to ones focused around scalar-tensor theories \cite{Creminelli:2008wc,Gubitosi:2012hu,Bloomfield:2012ff,Gleyzes:2013ooa,Gleyzes:2014rba}. While we would like to be completely general, as a first step we will focus on a large, subclass of 
scalar-tensor theories described by the Horndeski action \cite{Horndeski:1974wa,Deffayet:2009wt}. This will allow us to see how future constraints will improve our knowledge of what is currently a well understood swathe of model space.

The Horndeski action encompasses all scalar-tensor theories (with one scalar field) that have second order equations of motion. It includes quintessence, K-essence, Jordan-Brans-Dicke \cite{Brans:1961sx} and its variants, and generalized covariant Galileons. Substantial effort has already gone into studying the cosmological dynamics of these theories \cite{DeFelice:2011hq} as well as attempts at constraining them \cite{Avilez:2013dxa,Ade:2015rim,Bellini:2015xja} or forecasting future constraints from specific future surveys \cite{Piazza:2013pua,Gleyzes:2015rua,Leung:2016xli}. In this paper we will  include key surveys that are planned to come online in the next decade or so. Considering this subclass of models will allow us to better understand the role that priors play in forecasting constraints and, in particular, how priors on the gravitational theories will interact with the priors on the usual cosmological parameters. It will also allow us to compare directly with constraints on smaller scales, i.e. 
with the fabled precision constraints from the Solar System, and binary systems \cite{Berti:2015itd}.

This paper is structured as follows. In Section \ref{sec:horndeski} we present Horndeski's formulation of scalar tensor theories and identify the class of theories we will be considering in our analysis. We take particular care in identifying three different levels of parametrizations which we will be forecasting for. In Section \ref{sec:surveys} we list the surveys we will be considering and the forecasting method we will be using; our focus is primarily on tomographic surveys supplemented with intensity mapping experiments. We do briefly consider the added value of spectroscopic measurements of the baryon acoustic oscillation scale (BAO) and of the growth rate via redshift space distortions. In Section \ref{sec:results} we present our results, taking particular care to identify the optimal combination of surveys, identifying the degeneracies between parameters and contrasting the future with the current state of play. In Section \ref{sec:discussion} we discuss our findings.

\section{Horndeski's formulation of Scalar-Tensor theories}\label{sec:horndeski}
A general scalar-tensor action, constructed from a metric, $g_{\mu\nu}$ and scalar field, $\phi$ was proposed in \cite{Horndeski:1974wa,Deffayet:2009wt} and takes the following form:
\begin{eqnarray}\label{eq:L_horndeski}
S=\int d^4x \sqrt{-g}\left\{\sum_{i=2}^5{\cal L}_i[\phi,g_{\mu\nu}]+{\cal L}_M[g_{\mu\nu},{\varphi}]\right\},
\end{eqnarray}
where ${\cal L}_M$ is the minimally coupled matter action ($\varphi$ represents a general "matter" field) and the building blocks of the scalar field lagrangian are
\begin{eqnarray}
{\cal L}_2&=& K ,  \nonumber \\
{\cal L}_3&=&  -G_3 \Box\phi , \nonumber \\
{\cal L}_4&=&   G_4R+G_{4X}\left\{(\Box \phi)^2-\nabla_\mu\nabla_\nu\phi \nabla^\mu\nabla^\nu\phi\right\}  , \nonumber \\
{\cal L}_5&=& G_5G_{\mu\nu}\nabla^\mu\nabla^\nu\phi
-\frac{1}{6}G_{5X}\big\{ (\nabla\phi)^3
-3\nabla^\mu\nabla^\nu\phi\nabla_\mu\nabla_\nu\phi\Box\phi 
 \nonumber \\ & & 
+2\nabla^\nu\nabla_\mu\phi \nabla^\alpha\nabla_\nu\phi\nabla^\mu\nabla_\alpha \phi
\big\}   \,.
\end{eqnarray}
We have that $K$ and $G_A$ are functions of $\phi$ and $X\equiv-\nabla^\nu\phi\nabla_\nu\phi/2$, and the subscripts $X$ and $\phi$ denote derivatives. The four functions, $K$ and $G_A$ completely characterize this class of theories. It is possible to extend this action by including extra terms (to generate what has been dubbed the ``beyond Horndeski'' action \cite{Zumalacarregui:2013pma,Gleyzes:2014dya,Gleyzes:2014qga}) ``extended scalar-tensor theories'' \cite{Langlois:2015cwa,Crisostomi:2016czh,BenAchour:2016fzp} or non-minimal coupling to matter \cite{Gleyzes:2015pma}.

From the point of view of cosmology, one can greatly reduce the number of degrees of freedom one needs to focus on to the homogeneous mode of the metric, $ds^2=-dt^2+a^2(t)d{\vec x}^2$, and scalar field, ${\bar \phi}(t)$, and their linear perturbations. In \cite{Bellini:2014fua} it was shown that, on top of a convenient choice of the background expansion history (described by the effective equation of state, $w$,  of the Horndeski field), one can completely characterize the evolution of linear perturbations in Horndeski cosmologies in terms of a set of free functions of time given by
\begin{eqnarray}
M^2_*&\equiv&2\left(G_4-2XG_{4X}+XG_{5\phi}-{\dot \phi}HXG_{5X}\right) , \nonumber \\
HM^2_*\alpha_M&\equiv&\frac{d}{dt}M^2_* , \nonumber \\
H^2M^2_*\alpha_K&\equiv&2X\left(K_X+2XK_{XX}-2G_{3\phi}-2XG_{3\phi X}\right) \nonumber \\ & &
+12\dot{\phi}XH\left(G_{3X}+XG_{3XX}-3G_{4\phi X}-2XG_{4\phi XX}\right) \nonumber \\ & &
+12XH^2\left(G_{4X}+8XG_{4XX}+4X^2G_{4XXX}\right)\nonumber \\ & &
-12XH^2\left(G_{5X}+5XG_{5\phi X}+2X^2G_{5\phi XX}\right)\nonumber \\ & &
+14\dot{\phi}H^3\left(3G_{5X}+7XG_{5XX}+2X^2G_{5XXX}\right)\nonumber , \\
HM^2_*\alpha_B&\equiv&2\dot{\phi}\left(XG_{3X}-G_{4\phi}-2XG_{4\phi X}\right) \nonumber \\ & &
+8XH\left(G_{4X}+2XG_{4XX}-G_{5\phi}-XG_{5\phi X}\right)  \nonumber \\ & &
+2\dot{\phi}XH^2\left(3G_{5X}+2XG_{5XX}\right) \nonumber , \\ 
M^2_*\alpha_T&\equiv&2X\left[2G_{4X}-2G_{5\phi}-\left(\ddot{\phi}-\dot{\phi}H\right)G_{5X}\right] \,.
\end{eqnarray}
Each of these functions can be tied to its own underlying physical aspect of scalar tensor theories: $M^2_*$ and $\alpha_M$ are related to time variations in the background Newton's constant, $\alpha_K$ (dubbed "kineticity") generalizes the canonical kinetic term of simple DE models while $\alpha_B$ (dubbed "braiding") quantifies kinetic mixing between $\phi$ and the scalar perturbations of the metric. Finally, $\alpha_T$ is associated to modifications to the speed of propagation of tensor modes, but it is also responsible for anisotropic stress in the scalar sector, i.e.\ $\gamma$ defined in Eq.\ \ref{eq:qs-parameters}.\footnote{Both $\alpha_M$ and $\alpha_T$ affect the propagation of gravitational waves and can be constrained by the period decay in binary systems \cite{Jimenez:2015bwa} and future detections from systems with electromagnetic or neutrino counterparts \cite{Lombriser:2015sxa,Bettoni:2016mij}.}

Furthermore, one can infer from this parametrization, emergent scales which define how perturbations evolve. As orginally discussed in \cite{Bellini:2014fua} and further elaborated in \cite{Bellini:2015xja}, the "braiding scale", 
\begin{eqnarray}\label{eq:k_braiding}
k_B =\frac{D}{\alpha_B^2}\left[\left(1-\Omega_M\right)(1+w)+2(\alpha_M-\alpha_T)\right]+\frac{9}{2}\Omega_M,
\end{eqnarray}
(where the time-varying Planck mass has been absorbed  into the definition of $\Omega_M$ and $D\equiv\alpha_K+\frac{3}{2}\alpha^2_B $) determines the scale dependence in the growth factor.

The set of $\alpha$s are a remarkably compact and efficient way of characterizing this large sub-class of scalar tensor theories. On the one hand, it greatly restricts the possible structure and evolution of non-GR cosmologies; this finite-dimensional subspace of theories can be easily parametrized by a handful of numbers characterizing their time evolution. One obvious, simplified, approach is to associate the onset of modifications to GR with the emergence of dark energy; indeed one of the motivations to consider such theories is a possible explanation for the late-time acceleration of the Universe. Thus a
possible parametrization is
\begin{eqnarray}\label{eq:de_alphas}
\alpha_X=f\left(\frac{\Omega_{DE}}{\Omega_{DE0}}\right), \label{param1}
\end{eqnarray}
where $\Omega_{DE}$ is the fractional energy density in dark energy (and the additional subscript "$0$" denotes its value today) and $f(x)$ is a function such that $f\rightarrow0$ as $x\rightarrow 0$.
To describe gravity at low redshift we can Taylor expand this function and, for example, simply keep the leading order term, so that $f\simeq f'(0)x$. A recent analysis of this class of theories using this parametrization was undertaken in \cite{Bellini:2015xja} where the uncertainty, from current data, on $\alpha_B$ and $\alpha_M$ was found to be ${\cal O}(1)$.

As stated, this is a simplified approach and does not necessarily capture the correct behaviour of the $\alpha$s arising in Horndeski theory \cite{Linder:2015rcz}. Simply put, if one were to start with the Horndeski action for the background evolution, choosing a physical range of initial conditions and free functions, one would then derive a set of theoretical priors for the time evolution of the $\alpha$s. One would then construct a parametrization of this prior space of $\alpha$s to be used in an accurate forecast analysis. This is analogous to what was shown for the case of the simple ($w_0$, $w_a$) parametrization of dark energy where a reasonably tight correlation between the two parameters exists already at the level of the priors \cite{Marsh:2014xoa}. Nevertheless, our simple parametrization does allow us to get a sense of how constraining future data sets will be in comparison with
current data. In addition, this parametrization describes exactly the behavior on their attractor of simple shift-symmetric models as the imperfect fluid \cite{Pujolas:2011he}, and approximately more complicated models as the best fit of covariant galileons without a cosmological constant \cite{Barreira:2014jha,Renk:2016olm}.

Within this restricted class of models, we can also map directly onto a set of parameters which are of particular use in the quasi-static regime, when $c_s ak/H\gg 1$ (where $k$ is the wavenumber of a given perturbation, $a$ is the scale factor, $H$ is the Hubble rate and $c_s$ is the sound speed of the additional degree of freedom). The linear perturbation equations in the Newtonian gauge greatly simplify to
\begin{equation}\label{eq:qs-parameters}
-2 \frac{k^2}{a^2} \Phi = G_0\mu {\bar \rho}\delta,\hspace{12pt}
\frac{\Psi}{\Phi}=\gamma,
\end{equation}
where $\Phi$ and $\Psi$ are the gravitational potentials (using the metric defined in \cite{Bellini:2014fua}), ${\bar \rho}$ is the background energy density of non-relativistic matter with perturbations $\delta$. It has been shown that the two quasi-static parameters - the modified Newton's constant, $G_{\rm eff}=\mu G_0$ and the gravitational slip, $\gamma$ - can be expressed as rational functions of $k$ with time dependence uniquely determined by the $\alpha$s. Their expressions can be found in Appendix \ref{app:quasi-static}.
While we won't actually forecast constraints on ($\mu$,$\gamma$) in this paper, it is useful to assess what are current constraints. The most complete analysis of quasi-static parameters can be found in \cite{Ade:2015rim} where
a set of somewhat paradoxical constraints were found. On the one hand, the current uncertainty in
($\mu$,$\gamma$) is of ${\cal O}(1)$, although these parameters do suffer from a strong degeneracy; if we look at the direction orthogonal to that degeneracy, we can bring down the uncertainty
to ${\cal O}(10^{-1})$. On the other hand, the use of these quasi-static parameters provided the first clear indication of a deviation from GR. While it is reasonably robust to different permutations of data sets, the authors of \cite{Ade:2015rim} have opted to claim that this is a signal of systematic effects in the data. Future datasets, such as the ones we consider should unambiguously resolve this issue.

So far we have explained how we will use the Horndeski parametrization to explore a broad sub-class of theories. We can also do the converse and consider a very narrow sub-class of well-known theories that are embedded in Horndeski. We will thus also focus on Jordan-Brans-Dicke (JBD) theories \cite{Brans:1961sx} which are described in terms of
\begin{eqnarray}
S=\int d^4x \sqrt{-g}\left[\phi R-\frac{\omega_{\rm BD}}\phi\nabla_\mu\phi\nabla^\mu\phi+V+{\cal L}_M[g_{\mu\nu},{\varphi}]\right\}\,. \nonumber \\
\end{eqnarray}
When $V=0$, these theories are completely described in terms of one parameter, $\omega_{\rm BD}$, and GR is recovered when $\omega_{\rm BD}\rightarrow \infty$.
This theory is the workhorse of modern tests of GR on a wide range of scales and allows us to compare the potential of future cosmological surveys with other astrophysical tests. This theory will
trace out a particular slice in the space of $\alpha$s.

JBD theories are exquisitely constrained. The current tighest constraint on $\omega_{\rm BD}$ is on astrophysical scales from the Shapiro time delay \cite{Bertotti:2003rm}: $\omega_{\rm BD} >4\times10^{4}$. The most up-to-date cosmological constraints using the Planck data are in \cite{Avilez:2013dxa} where it was found that $\omega_{\rm BD} >6\times 10^2$. See also \cite{Umilta:2015cta,Ballardini:2016cvy}). Naturally, the question arises whether it is possible to get cosmological constraints on $\omega_{\rm BD}$ comparable to those found on astrophysical scales.

Many alternative theories of gravity are equipped with screening mechanisms, non-linear effects by virtue of which the departures from GR fade in short scales or high-density environments \cite{Joyce:2014kja}. 
This feature helps reconcile such theories with the stringent astrophysical and Solar System tests while leaving room for detectable signatures in the large-scale structure of the Universe. However, screening typically kicks in on small cosmological scales, suppressing the modified gravity effects relative to the linear prediction (see \cite{2013JCAP...11..012L,Barreira:2013eea,2013JCAP...05..023L}and for an extensive cross comparison, see \cite{Winther:2015wla}).
This feature of modified gravity has been studied in a model-by-model basis using non-linear techniques, but has been often ignored in forecasts for future experiments.
We will use linear cosmological perturbation theory and hence the screening effects have to be included in a phenomenological fashion. We will model the small-scale recovery of GR through a scale dependence of the $\alpha$s
\begin{equation}\label{eq:vainshtein_cutoff}
 \alpha_X(t)\to\alpha_X(t,k)=\alpha_X(t)S\left(\frac{k}{k_V}\right),
\end{equation}
where $X=(M,B,K,T)$ and such that $S(x\ll1)\approx 1$ and $S(x\gg1)=0$. We will model the screening using a Gaussian function $S=\exp \big(-\frac{1}{2}\left({k}/{k_V}\right)^2)$ with a fiducial value $k_V=0.1 h/$Mpc, in agreement with simulations of Vainshtein-screened models (cf. Fig. 4 of \cite{Barreira:2013eea}). This is a minimal, one-parameter prescription. More sophisticated prescriptions can be added as needed based on model-specific studies. We note that this or other prescription needs to be introduced whenever screened scales are included in the analysis. If they can be properly modeled, non-linear scales contribute greatly to constraint other cosmological parameters, but not accounting for screening can largely overestimate the surveys capacity to test gravity.

The situation with regards to baryonic physics is even more open: while there have been attempts at understanding the impact of, for example, AGN feedback on specific models \cite{2013MNRAS.436..348P}, a complete understanding is still lacking. In order to account for the effect of baryonic effects to some extent, we have used the model proposed by \cite{2015JCAP...12..049S}, determined by two quantities, $M_c$ and $\eta_b$, parametrising the mass dependence of the halo gas fraction and the ejected gas radius. For these we use the fiducial values $M_c=1.2\times10^{14}\,M_\odot/h$ and $\eta_b=0.5$. In Section \ref{ssec:smallscales} we will study the impact of the uncertainties on both screening mechanisms and baryonic physics by marginalising over the values of these parameters as well as $k_V$.

\section{Stage IV cosmological surveys}\label{sec:surveys}
In the next decade, a number of astronomical facilities will cover a large portion of the sky visible from the Southern Hemisphere in multiple wavelengths and with almost perfect angular overlap. We will thus have at our disposal a large volume of the Universe where cosmological structures will have been covered by a large variety of observational probes. In this section we describe the forecasting method that we use to predict the combined constraining power of a representative subset of these experiments, the nature of the main cosmological probes pursued by them, and the main advantages of combining probes.

\subsection{Forecasting formalism}\label{ssec:fisher}
Each probe $a$ considered in this paper can be represented by a set of $N_{\rm maps}^a$ sky maps fully described by their spherical harmonic coefficients $a_{\ell m}^{(a,i)}$ ($i\in[1,N_{\rm maps}^a]$). These may correspond, for instance, to the perturbations in the number density of galaxies in a set of redshift bins or to the temperature and polarization perturbations of the CMB. Each sky map is labelled by a combined index $(a,i)\rightarrow A$, and to first approximation we will assume that the anisotropies $a_{\ell m}^A$ are Gaussianly distributed, such that the combined likelihood of the observed maps is given by
\begin{equation}
  -2\log p({\bf a}_{\ell m})=
  \sum_\ell \sum_{m=-\ell}^\ell \left\{{\bf a}^T_{\ell m}{\sf C}_\ell{\bf a}_{\ell m}+
  \log[{\rm det}(2\pi\,{\sf C}_\ell)]\right\}.
\end{equation}
Here we have grouped all the sky maps into a vector ${\bf a}_{\ell m}$, and we have defined the power spectrum matrix ${\sf C}_\ell$ as the covariance of this vector:
\begin{equation}
  \left\langle{\bf a}_{\ell m}\,{\bf a}^\dag_{\ell'm'}\right\rangle=
  \delta_{\ell\ell'}\delta_{mm'}{\sf C}_\ell.
\end{equation}

The information on cosmological parameters is encoded in the power spectrum. Expanding this likelihood around the maximum we can find the now usual expression for the Fisher matrix ${\sf F}$, describing the inverse of the covariance matrix of a set of parameters $\mathbf{\theta}$:
\begin{equation}\label{eq:fisher}
 {\sf F}_{\mu\nu}=\sum_{\ell=2}^{\ell_{\rm max}} f_{\rm sky}(\ell+1/2)\,
 {\rm Tr}\left(\partial_\mu{\sf C}_\ell\,{\sf C}_\ell^{-1}\partial_\nu{\sf C}_\ell{\sf C}_\ell^{-1}\right),
\end{equation}
where $f_{\rm sky}$ is the sky fraction covered by the considered probes, and $\partial_\mu$ implies differentiation with respect to the parameter $\theta_\mu$. The maximum multipole $\ell_{\rm max}$ included in the constraints is a map-dependent choice, corresponding to the smallest scale for which sensible information can be extracted, and can be limited by noise and observational or theoretical systematics.

We will assume that the observed anisotropies contain both a cosmological signal and a noise component, ${\bf a}_{\ell m}={\bf s}_{\ell m}+{\bf n}_{\ell m}$, and that both components are uncorrelated. Thus the power spectrum can also be decomposed into two components, ${\sf C}_\ell={\sf C}^S_\ell+{\sf C}^N_\ell=\langle{\bf s}_{\ell m}{\bf s}^\dag_{\ell m}\rangle+\langle{\bf n}_{\ell m}{\bf n}^\dag_{\ell m}\rangle$. The form of the signal and noise components for each probe will be described below.

The partial derivatives in Eq. \ref{eq:fisher} were computed using finite central differences:
\begin{equation}
 \partial_\mu f(\mathbf{\theta})=\frac{f(\theta_\mu+\delta\theta_\mu)-f(\theta_\mu-\delta\theta_\mu)}{2\delta\theta_\mu},
\end{equation}
where the optimal value for the intervals $\delta\theta_\mu$ was found by iteratively halving the initial guesses until convergent results were found beyond the second significant digit. All power spectra were computed using a modified version of \hiclass\footnote{\url{www.hiclass-code.net}} \cite{Zumalacarregui:2016pph}, a code based on the Cosmic Linear Anisotropy Solving System \cite{Blas:2011rf}. \hiclass\, computes the evolution of linear cosmological observables without assuming the quasi-static approximation and includes relativistic corrections to galaxy clustering \cite{DiDio:2013bqa}. This ensures the validity of the computation on scales larger than the scalar field sound horizon \cite{Sawicki:2015zya} and all the way to ultra-large scales. In order to retain the correct large-scale shape, all power spectra were computed without adopting the Limber approximation. \hiclass\, was run adding a constant value to the kineticity $\alpha_K$, i.e.\ $10^{-4}$, to protect the computation against numerical singularities in the evolution of the perturbations that happen when the kinetic term $D\rightarrow0$. In addition, \hiclass\, uses a set of precision parameters which is improved w.r.t.\ to the ones in CLASS \cite{Blas:2011rf}.

\subsection{Cosmological surveys}
\subsubsection{Stage-4 CMB}
The current CMB datasets, consisting of a combination of large-scale, full-sky experiments such as WMAP \cite{2013ApJS..208...20B} and Planck \cite{2014A&A...571A..16P}, and high-resolution ground-based observatories (e.g. ACTPol \cite{2014JCAP...10..007N}, SPT-Pol \cite{2015ApJ...807..151K}, are currently being upgraded through Stage 3 (S3) ground-based experiments (e.g. AdvACT \cite{2016JLTP..184..772H} and SPT-3G \cite{2014SPIE.9153E..1PB}) with a larger numbers of detectors, multiple frequency channels, and covering larger sky fractions.

S3 experiments will eventually be superseded by a Stage 4 (S4) experiment, likely to be built by combining the observing power of different ground-based facilities, with similar potential for wide sky coverage and significantly lower noise levels. It is expected that S4 will cover $\sim\!40\%$ of the sky, with a reduced noise level of around $\sigma_T=1\uKam$ in temperature \cite{Abazajian:2016yjj}. 

S4 will measure three main types of anisotropies: the primordial CMB perturbations in temperature and polarization ($a_{\ell m}^T,a_{\ell m}^E$\footnote{Although both the $E$- and $B$-modes contribute to the total polarized anisotropies, here we will set the primordial tensor perturbations (and hence the $B$-modes) to zero. The contribution of the lensing $B$-modes is therefore fully accounted for by considering the reconstructed CMB lensing convergence}), as well as the reconstructed CMB lensing convergence $a_{\ell m}^\kappa$. For the CMB anisotropies, the noise power spectrum is determined by two regimes. On small scales, the measurements are limited by the beam size of S4, which here we assume to be Gaussian with a width $\theta_{\rm FWHM}=3'$. In this regime the noise can be approximated as being white (before noise deconvolution \cite{2016arXiv161002360L}), and thus given by:
\begin{equation}
 N^{T,E}_\ell=\sigma_{T,E}^2\,\exp[\ell(\ell+1)\theta_{\rm FWHM}^2/(8\ln2)],
\end{equation}
with $\sigma_{T,E}^2$ in units of $\mu{\rm K}^2\,{\rm sr}$ (we assume $\sigma_E=\sigma_T\sqrt{2}$). On the other hand, ground-based facilities such as S4 are limited on large scales by a number of statistical and systematic uncertainties, such as the effect of atmospheric noise or ground pickup. For this reason we assume that S4 will not be able to probe the CMB anisotropies on scales $\ell<30$, and in those multipoles we assume noise levels corresponding to Planck as given in \cite{2014A&A...571A..16P}. Furthermore, we set the maximum multipoles for these probes at $\ell_{\rm max}=3000$ in temperature and $\ell_{\rm max}=5000$ in polarization\footnote{The different small-scale cutoff in temperature and polarization is motivated by the contamination from extragalactic foregrounds on small scales in the temperature power spectrum.}.

In order to compute the noise power spectrum for the lensing convergence, we assume a reconstruction process based on quadratic combinations of the lensed CMB maps \cite{2006PhR...429....1L}, and estimate the reconstruction noise as detailed in \cite{2002ApJ...574..566H}. In doing this we assume a minimum-variance noise achieved by combining the $TT$, $TE$, $TB$, $EE$ and $EB$ estimators, using only the multipole range $30<\ell<3000$.

It is worth noting that our formalism accounts for the non-zero correlation between overlapping CMB and large-scale structure measurements, both those caused by CMB lensing and by the late-time integrated Sachs-Wolfe effect.

\subsubsection{The Large Synoptic Survey Telescope}
Photometric redshift (photo-$z$) surveys have been proposed as a practical means to achieve simultaneously dense and deep galaxy catalogs. In this approach, the redshift of each galaxy is inferred from its flux in a small number of wide frequency bands, and although the corresponding redshift uncertainties prevent any efficient measurement of the radial clustering pattern, the large number density of tracers attainable by these surveys makes them ideal for weak lensing studies as well as multi-tracer analyses.

Although large-scale photometric surveys, such as the Dark Energy Survey \cite{2005astro.ph.10346T}, are already underway, we will focus our discussion on the Large Synoptic Survey Telescope (LSST), a Stage-IV experiment aiming at covering $\sim\!20,000\,{\rm deg}^2$ on the sky to with a magnitude limit $r\sim27$. We will concentrate here on two of the main cosmological science cases covered by the LSST: galaxy clustering and cosmic shear.

\paragraph{Galaxy clustering.} The main cosmological observable in galaxy clustering studies is the shape of the power spectrum or two-point correlation function of the galaxy number density. In the case of LSST we will focus on a tomographic approach in which the galaxy sample is first separated into a number of bins of photo-$z$, and all auto- and cross-correlations between these bins are analyzed simultaneously.

Since the galaxy number density is known to be a biased tracer of the matter density field ($\delta_g\simeq b\,\delta$), the uncertainties in the redshift- and possibly scale-dependent bias $b$ prevent us from using the amplitude of perturbations in the galaxy overdensity to make an independent measurement of the growth of structure. As a result, most of the information in the angular galaxy-galaxy power spectrum resides in robust features such as the angular BAO scale. Other ``secondary'' contributions to the power spectrum, such as that of redshift-space distortions can also be used to make a sub-optimal measurement of the growth rate and, although sub-dominant, a number of relativistic contributions to the clustering pattern, such as that of lensing magnification \cite{Yoo:2009au,2011PhRvD..84d3516C,2011PhRvD..84f3505B} could potentially contain relevant information about deviations from GR.  Further details can be found in \cite{Alonso:2015uua}. These expressions are general for minimally coupled 
theories of gravity in the linear regime \cite{Renk:2016olm}.

Under the assumption that galaxies are a Poisson sample of the underlying biased density field, we can model the noise power spectrum as a white component with an amplitude of $1/N_\Omega$, where $N_\Omega$ is the angular number density of sources in each redshift bin in units of ${\rm sr}^{-2}$. As was done in \cite{Alonso:2015sfa}, we separate the clustering sample into two sub-samples of ``red'' (early-type, ellipticals, high-bias) and ``blue'' (late-type, disks, low-bias) galaxies. The specific models for the power spectrum, photo-$z$ uncertainties, nuisance bias parameters and sample definitions are described in detail in \cite{Alonso:2015uua}

Although a linear, scale-independent bias parameter is in most cases a good approximation to the relation between the galaxy and matter power spectra on large scales, on small non-linear scales it becomes necessary to resort to more sophisticated models, which often prevents the use of the small-scale galaxy power spectrum for cosmology. In order to avoid these complications we define a minimum angular scale for each redshift bin down to which the corresponding map is used. At the mean redshift of each bin $\bar{z}$ we start by defining a threshold comoving scale $k_{\rm max}$ by requiring that the variance of the matter overdensity on larger scales be below a given threshold $\sigma_{\rm thr}^2$, i.e:
\begin{equation}\label{eq:sigma_thr}
  \sigma_{\rm thr}^2=\frac{1}{2\pi^2}\int_0^{k_{\rm max}}dk\,k^2\,P(k,\bar{z}).
\end{equation}
We then translate this comoving scale into an angular multipole $\ell_{\rm max}=\chi(\bar{z})\,k_{\rm max}$. For our fiducial constraints we used a threshold variance of $\sigma_{\rm thr}=0.5$. This value corresponds to a cutoff scale $k_{\rm max}\sim0.1 h\,{\rm Mpc}^{-1}$ at $z=0$, which is a conservative estimate of the scales up to which a reliable estimate of the covariance matrix of the matter power spectrum can be obtained using perturbation theory \cite{2017MNRAS.466..780M}.

\paragraph{Cosmic shear.} The observed projected shapes of galaxies are distorted due to the weak gravitational lensing of the photons they emit caused by the intervening matter perturbations. Thus it is possible to probe those perturbations by analysing the  correlated galaxy ellipticities. The constraining power of weak leansing can be summarized into the power spectrum of the traceless part of the cosmic shear tensor computed for galaxies lying in a set of photo-$z$ bins (labelled here by an index $i$):
\begin{align}
  \gamma_i(\nv)&\equiv\int_0^{\chi_H} d\chi'\,
                  W^i_\gamma(\chi)\eth\eth(\Phi+\Psi)(\chi,\nv),\\\nonumber
  W^i_\gamma(\chi)&\equiv  
  \int_\chi^{\chi_H} d\chi' \frac{dp_i}{d\chi'}\left(\frac{\chi'-\chi}{\chi\chi'}\right),
\end{align}
where $dp_i/d\chi$ is the selection function of the redshift bin, and $\eth$ is the spin-raising differential operator defined by their action on a spin-$s$ function $f_s$:
\begin{equation}
 \eth f_s(\nv) = -(\sin\theta)^s(\partial_\theta+i\partial_\phi/\sin\theta)
 (\sin\theta)^{-s}f_s(\nv).
\end{equation}
We see that, at any particular redshift, weak lensing traces the density perturbations integrated along the line of sight down to the observer weighed by the lensing kernel. Thus, good redshift precision is not necessary for this probe, since the large support of the lensing kernel erases all structure along the line of sight, making photometric redshift surveys ideal for this task.

Since the shear tensor is effectively estimated by averaging over the ellipticities of all galaxies laying in a given pixel, the noise in this estimate is directly proportional to the variance of the intrinsic galaxy ellipticities, and inversely proportional to the angular galaxy number density. We thus use a white noise model in which the noise power spectrum for the $i$-th redshift bin is given by
$N_\ell^i=\sigma_\gamma^2/N_\Omega^i$, where $\sigma_\gamma^2$ is the per-component dispersion in the intrinsic galaxy ellipticities (including measurement noise), for which we use $\sigma_\gamma=0.3$ \cite{2009arXiv0912.0201L}.

Since weak lensing directly probes the matter perturbations, theoretical uncertainties on non-linear scales are in principle far less cumbersome than in the case of galaxy clustering. However, even using numerical simulations and emulators \cite{2010ApJ...715..104H}, the uncertainties in the modeling of baryonic effects prevent an accurate description of the matter power spectrum on very small scales \cite{2008ApJ...672...19R,2011MNRAS.415.3649V,2014MNRAS.440.2997V,2016MNRAS.461L..11H,2015MNRAS.450.1212H,2016MNRAS.461L..11H}. For these reasons, for weak lensing we use a small-scale cutoff $\ell_{\rm max}=2000$, corresponding to $k_{\rm max}\sim1$ at $z=1$ (note that we will use a different prescription in Section \ref{ssec:smallscales}, with $\ell_{\rm max}$ defined as above for galaxy clustering). 

Cosmic shear is not free from systematic uncertainties: shape measurement uncertainties are known to affect the broadband shape of the lensing power spectrum, and multiplicative and additive bias parameters have been used to model the effects in the first cosmological lensing analyses \cite{2013MNRAS.432.2433H,2015arXiv150705552T,2016arXiv160605338H}. Furthermore, intrinsic galaxy orientations are known to be correlated with each other, and there are hints that the local tidal field (i.e. the Hessian of the gravitational potential) is responsible for these intrinsic alignments, at least in the case of elliptical galaxies. This effect can also be modelled at the cost of introducing extra nuisance parameters. This work aims at forecasting the best achievable constraints on scalar-tensor theories, and therefore we will ignore shape-measurement systematics, assuming that tight priors on the corresponding nuisance parameters can be found below the science requirements of LSST. We have also neglected the systematic uncertainties associated with the use of photometric redshifts, which would also affect galaxy clustering. We have, however, included the effect of intrinsic alignments following the so-called non-linear alignment model \cite{2004PhRvD..70f3526H}, marginalising over the alignment amplitude at $z\in\{0.5,1.0,1.5,2.0\}$. We defer a more thorough study of the effect of systematic uncertainties on modified gravity constraints for future work.

\subsubsection{The Square Kilometre Array}
A wide-area radioastronomy facility such as the Square Kilometre Array (SKA) offers a large variety of  unique as well as synergistic cosmological science cases, which have recently received a lot of attention in the literature \cite{2014MNRAS.442.2511F,2015ApJ...803...21B,2015aska.confE.145B,2015aska.confE.146K}. Three types of cosmological surveys can be carried out in this range of frequencies:
\begin{itemize}
 \item In an {\sl HI galaxy survey} \cite{2015MNRAS.450.2251Y}, individual sources are spatially resolved, and their 21cm, neutral hydrogen detected with significantly high signal-to-noise, thus producing a spectroscopic catalog of galaxies. The low intensity of this line makes it very challenging to reach high number densities and redshifts with this technique, and it is expected that only a hypothetical ``Phase-2'' of the SKA would be able to produce a HI catalog competitive with planned Stage-IV surveys such as Euclid \cite{Laureijs:2011gra}.
 \item Dropping the requirement of resolving the HI line, a {\sl continuum survey} \cite{2015aska.confE..18J} simply integrates the total radio flux of all sources in a wide frequency band. This allows the detection of much fainter sources, and a continuum survey with a flux limit of $5\mu{\rm Jy}$ would be able to cover a wide range of redshifts ($z\lesssim4$). The main drawback of this technique is the lack of redshift estimates for the detected sources, which in the best-case scenario makes them dependent on external datasets and optical follow-up, and in the worst renders them useless for cosmological studies. It is also worth noting that, given the good control over the point-spread function achievable with a radio interferometer, radio weak lensing surveys from continuum catalogs could potentially be more robust against shape-measurement systematics than their optical counterparts \cite{2016MNRAS.tmp.1475H}.
 \item Conversely, the novel technique of {\sl intensity mapping} \cite{2015ApJ...803...21B} sacrifices angular resolution, avoiding the detection of individual sources by integrating the combined 21cm emission in wider angular scales, thus producing three-dimensional maps of the distribution of neutral hydrogen in the Universe with good radial resolution. Intensity mapping is therefore complementary to the use of photometric redshift surveys in the coverage of the $k_\perp-k_\parallel$ plane. 
\end{itemize}

Given its complementarity with the science cases covered by LSST, we will focus here on intensity mapping only. The cosmological observable in this case is the HI antenna temperature measured in a set of frequency bins, related to the corresponding redshift as $\nu=\nu_{\rm 21cm}/(1+z)$. We will assume that the observations will be done using the SKA1-MID as described in \cite{2015aska.confE..19S,Alonso:2015uua} we will dub this data set SKA-IM. We consider the frequency band 350-1050 MHz, dividing it into 200 frequency channels, corresponding to a comoving width of $\sim16\,{\rm Mpc}/h$. The models used to describe the signal and noise power spectrum of this observable, including the models for the HI bias and background temperature, are presented in detail in \cite{Alonso:2015uua}. We assume a 10,000-hour survey covering 40\% of the sky carried out with a set of 200 15 m antennas in single-dish observation mode with a system temperature of 25 K. The noise power spectrum is modelled as white (before beam de-convolution):
\begin{equation}
  N_\ell(\nu)=\sigma_N^2(\nu)\exp[\ell(\ell+1)\theta_{\rm FWHM}^2(\nu)/(8\ln2)],
\end{equation}
where the noise variance $\sigma_N$ and beam width $\theta_{\rm FWHM}$ are determined by the parameters listed above as described in \cite{Alonso:2015uua}. The beam size of the SKA ($\theta_{\rm FWHM}\sim2^\circ$ at $z\sim1$) is large enough that we do not need to impose a strict high-$\ell$ cut, since the measurements become noise-dominated well before the scale of non-linearities (in practice we impose a cut $\ell_{\rm max}=200$).

Since individual sources are not detected, the faint 21cm emission needs to be separated from the much brighter ($\sim5$ orders of magnitude) diffuse galactic and extragalactic foregrounds. Although, given the smooth frequency dependence of the foregrounds, it should be possible to isolate the cosmological signal based on its different spectral properties (see \cite{2014MNRAS.441.3271W,2015MNRAS.447..400A}), foreground residuals will necessarily dominate the measurements on large radial scales, and it is expected that foreground contamination coupled with instrumental mis-calibration will be the largest source of systematic uncertainties. Another cause of concern specific to single-dish observations is the effect of gain fluctuations in the time domain, which could be an important source of systematic uncertainties on large angular scales, although the effect may depend on the survey scanning strategy. As before, we will, for the most part, ignore these systematics in this work, in an attempt to present the 
best achievable constraints on scalar-tensor theories from future experiments.

\subsubsection{Spectroscopic surveys}\label{sssec:spec}
In our analysis we have not included constraints from wide spectroscopic such as DESI \cite{2011arXiv1106.1706S} or Euclid \cite{Laureijs:2011gra}. For this type of experiments, with good angular and radial resolution, the most efficient way to carry out Fisher forecasts is to use the Fourier coefficients of the galaxy overdensity $\delta_g(z,{\bf k})$ as an observable in a discrete set of redshift bins within which evolution effects are effectively frozen. In this formalism it is however not straightforward to account for inter-bin correlations \cite{2016arXiv160800458B} and correlations with overlapping lensing and CMB experiments, a key aspect of our analysis. On the other hand, the option of modelling spectroscopic observations as a set of angular maps at different redshift becomes computationally intractable without losing radial information. Furthermore, given that the experiments listed above cover the main science cases that a spectroscopic survey would be able to approach (geometric radial and angular BAO measurements and RSDs), we do not expect dramatic improvements in the final uncertainties due to the inclusion of spectroscopic data.

Nevertheless, and in order to estimate both the reach of future spectroscopic observations and the amount of information lost in our formalism, we have included constraints from an independent (i.e. uncorrelated) DESI-like spectroscopic survey using the expected uncertainties on the radial and angular BAO scales and the growth rate of structure estimated by \cite{Font-Ribera:2013rwa}. Besides simplicity, lack of correlation is an optimistic assumption that lets us evaluate how much the constraints may improve by the addition of this type of survey. We do not include these additional constraints as part of our fiducial forecasts, but discuss their relevance in Section \ref{ssec:overall}.

\section{Results}\label{sec:results}

The class of theories we are considering can, as we have seen, be parameterized in terms
of 5 free function of time, $w$, $\alpha_M$, $\alpha_K$, $\alpha_B$ and $\alpha_T$ (we fix the initial Planck mass, i.e.\ the integration constant needed to obtain $M_*^2$ from $\alpha_M$, to $1$). For the bulk of this analysis we assume that deviations from GR are intimately tied to the onset of accelerated expansion and hence, for now, we Taylor expand expression \ref{param1} and adopt a parametrization of the form:
\begin{equation}\label{eq:param_omega}
\alpha_X=b_X+c_X\,\frac{\Omega_{\rm DE}(z)}{\Omega_{\rm DE}(z=0)},
\end{equation}
where $\Omega_{\rm DE}$ is the fractional energy density in dark energy (or whatever
is responsible for the onset of accelerated expansion) which itself depends on $w$. For our fiducial constraints we fix the early terms to zero ($b_X=0$), and concentrate only on $c_X$. In a latter subsection we will consider $b_X$, as well as a different time dependence.

As a fiducial model we choose a point in the space of $c_X$ that is close enough to $\Lambda$ GR to be compatible within 1$\sigma$ given our most optimistic constraints (this is done to avoid the numerical singularities at $c_X=0$). The fiducial model we chose is $\{w=-1,\,c_K=0.1,\,c_B=0.05,\,c_M=-0.05,\,c_T=-0.05\}$. Beside these, we vary over the basic parameters of the flat $\Lambda$CDM model: the dark matter and baryon densities $\omega_c$ and $\omega_b$, the local expansion rate $h$, the amplitude and tilt of primordial scalar fluctuations ($A_s$, $n_s$) and the optical depth to reionization $\tau$. For these, we set their fiducial values to the best-fit cosmology of \cite{Ade:2015xua} (with $\tau=0.06$ as per the latest measurement of \cite{2016arXiv160502985P}). Furthermore, we consider a single massive neutrino with a mass of $60\,{\rm meV}$. When considering extended models with free early-time parameters, we set their fiducial values to $b_X=0$.

\subsection{Overall constraints.}\label{ssec:overall}

\begin{table*}
\centering
\begin{tabular}{l|c|c|c|c|c|c|c|c}
Case & $>\omega_{\rm BD},\,95\%\,{\rm C.L.}$ & $\sigma(c_B)$ & $\sigma(c_M)$ & $\sigma(c_T)$ & $\sigma(c_K)$ & $\sigma(w)$ & $\sigma(\sum m_\nu)\,[{\rm meV}]$ & ${\sl FoM}(c_B,c_M,c_T)$\\
\hline
S4                          & $2.9\times10^3$       & 0.796         & 0.746         & 1.26          & 4.9           & 0.112       & 71 & 1.3 \\
LSST                        & $1.2\times10^4$       & 0.193         & 0.089         & 0.205         & 8.8           & 0.016       & 45 & 61   \\
SKA1-IM                     & $9.5\times10^3$       & 13.3          & 6.0           & 8.6           & 106           & 0.018       & 74 & 1.0 \\
S4$+$LSST                   & $1.3\times10^4$       & 0.169         & 0.072         & 0.179         & 3.5           & 0.011       & 22 & 88   \\
S4$+$SKA1-IM                & $1.0\times10^4$       & 0.305         & 0.238         & 0.786         & 3.5           & 0.0085      & 23 & 9.0  \\
S4$+$LSST$+$SKA1-IM         & $1.7\times10^4$       & 0.161         & 0.070         & 0.151         & 3.1           & 0.0069      & 15 & 121  \\
S4$+$LSST$+$SKA1-IM$+$Spec. & $1.7\times10^4$       & 0.123         & 0.056         & 0.146         & 3.1           & 0.0061      & 13 & 143  \\
Best fit of \cite{Bellini:2015xja}   & N.A.                  & 0.063         & 0.076         & 0.201         & 4.23          & 0.0059      & 13 & N.A. \\
\hline
\end{tabular}
\caption{$1\sigma$ constraints on the Horndeski parameters $c_B$ and $c_M$, the dark-energy equation of state parameter $w$ and the sum of neutrino masses $\sum m_\nu$ for different combinations of experiments. The last column shows the constraints assuming a modified gravity fiducial model given by the best fit in \cite{Bellini:2015xja}, in which case $\Lambda$CDM would be ruled out by more than 7$\sigma$ from $c_M$ alone. The corresponding values for the figure-of-merit defined in Section \ref{sec:discussion} are shown in the last column.}
\label{table:sigmas_exp}
\end{table*}
\begin{figure*}[t]
      \centering
      \includegraphics[width=0.7\textwidth]{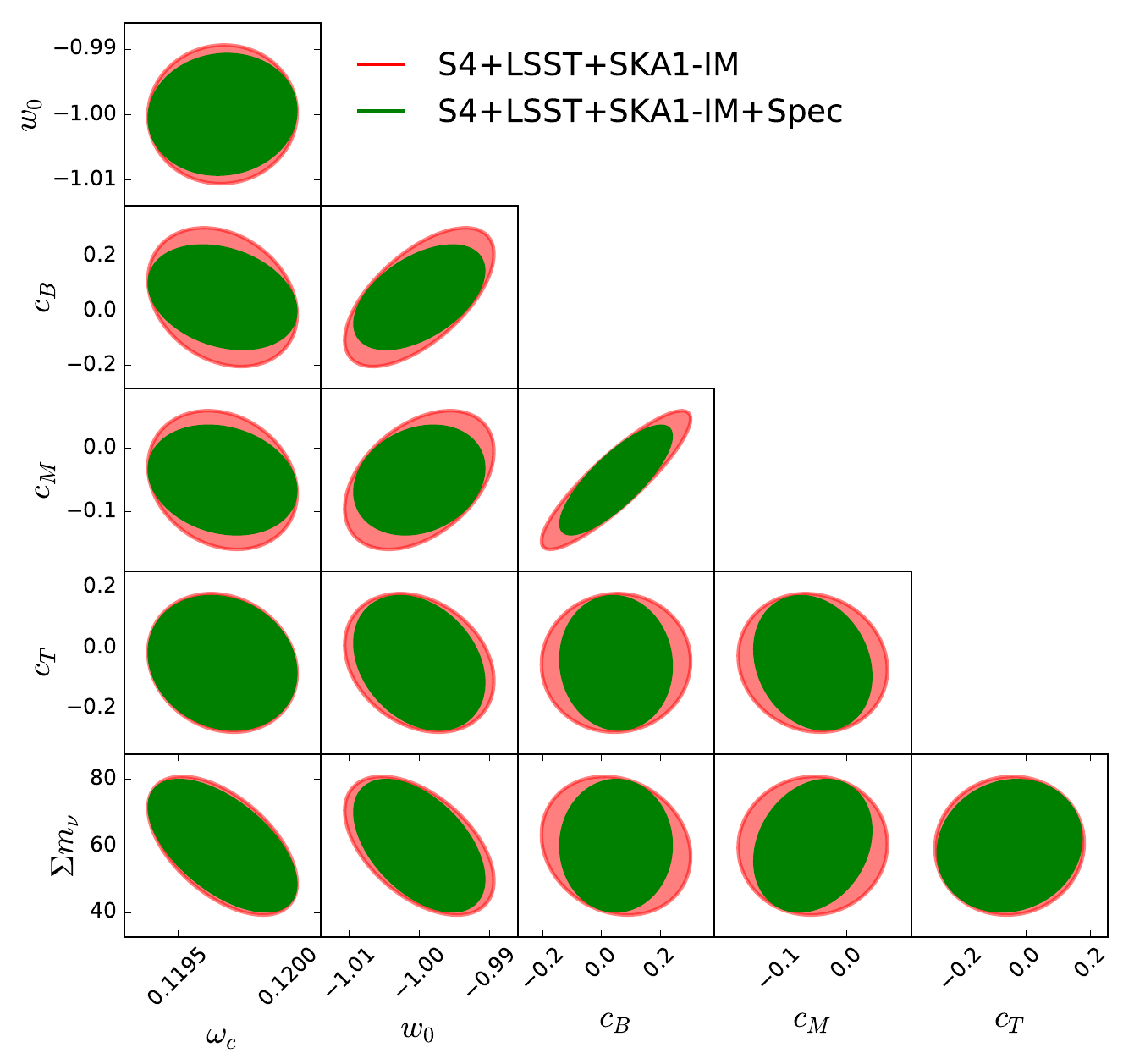}
      \caption{Cosmological constraints for the combination of CMB S4 with LSST (galaxy clustering and shear) and an intensity mapping experiment carried out by Phase-1 of the SKA (red ellipses). The green ellipses show the additional constraining power achievable by combining these observations with measurements of the BAO scale and the growth rate of structure carried out by an independent DESI-like experiment.}
      \label{fig:triangle_all}
      \vspace{-1em}
    \end{figure*}
We begin by considering the combination of our three main datasets (S4, LSST and SKA1-IM) to identiy the space of parameters on which the tightest constraints can be drawn. Table \ref{table:sigmas_exp} summarizes the forecast constraints on the most relevant parameters, and Fig \ref{fig:triangle_all} shows the covariance between them. The analysis of \cite{Bellini:2015xja}, gives us an idea of the overall structure of the constraints for fixed equation of state, $w=-1$,  with $c_M$ and $c_B$ being more tightly constrained than $c_K$ and $c_T$ (note that in \cite{Bellini:2015xja} the constraints on $c_T$ are heavily affected by theoretical priors, such as stability conditions). Our results show that, while the future generation of surveys we consider will also be able to pin down $c_T$ to a similar degree of precision, $c_K$ will remain a highly uncertain parameter. Fortunately, $c_K$ shows little or no degeneracy with any of the other Horndeski parameters, and therefore it can be marginalized over without 
degrading the constraints on $c_{B,M,T}$. This is explicitly shown in Table \ref{table:sigmas_marg}, which summarizes the degeneracies on these parameters for fixed or marginalized $c_K$. We see, however, that the uncertainties on $c_{B,M,T}$ grow between $10\%$ and $30\%$ when considering an evolving dark energy component with $w\neq-1$.

We can explore the full set of relevant degeneracies in Fig. \ref{fig:triangle_all}, which shows that $c_B$ and $c_M$ are tightly correlated along the direction $c_B\simeq 2.5\,c_M$. The dark energy equation of state also shows significant degeneracy with all the Horndeski parameters, especially $c_B$ and $c_M$, and our numerical results show also a non-negligible correlation with the fractional dark matter density $\omega_c$. These degeneracies are primarily driven by constraints on the growth of perturbations via the weak-lensing measurements in LSST (and less significantly through the RSD measurement from intensity mapping with SKA1). We can understand this from a brief analysis of the evolution equation for the growth rate, $f=d\ln\delta/d\ln a$:
\begin{eqnarray}
& \frac{df}{d\ln a}+q\,f+f^2=\frac{3}{2}\Omega_M \frac{\mu}{\gamma}, \label{f_eq}\\
& \mathrm{where}\quad\quad q(a)=\frac{1}{2}\left[1-3\,w(a) (1-\Omega_M(a)\right], \label{qdef}
\end{eqnarray}
$\delta$ is the matter density contrast and $\Omega_M$ is the fractional matter density. Note that $\Omega_M$ and $\mu/\gamma$ are time-dependent and, as $\mu/\gamma$ is a function of the $\alpha$s, we expect there to be a degenerate effect between these different parameters. The equation of state $w$ will affect $\Omega_M$ as well as $q$ leading to a further degeneracy. This also shows how crucial is to have precise distance measurements (specifically via the BAO) to be able to constrain this degeneracy.

While there is no significant degeneracy with the sum of neutrino masses (through their effect on $\Omega_M$, modifying Eq. \ref{qdef}), it is worth noting that, as has been found in other works, we predict a $\sim3\sigma$ measurement of the total neutrino mass (assuming the normal hierarchy lower bound) by combining S4 and low-redshift information from LSST ($\sigma(\sum m_\nu)=22\,{\rm meV}$).%
\footnote{This sensitivity could be increased to a $\sim5\sigma$ detection if an external, cosmic-variance-limited measurement of $\tau$ could be carried out ($\sigma(\sum m_\nu)=11\,{\rm meV}$) \cite{Allison:2015qca}.}
The lack of degeneracy is in contrast with previous studies based on specific models such as $f(R)$ \cite{Baldi:2013iza,Baldi:2016oce}. In this case the effect of neutrino masses (washing out perturbations on scales smaller than free-streaming scale) and modified gravity (enhanced growth due to the scalar force on scales shorter than the inverse mass of the field) can cancel partially. This cancellation is easier to achieve in restricted models (such as $f(R)$) due to the inter-dependences existing in the $\alpha$-functions (cf. \cite{Bellini:2014fua}). In more general models such as the one considered here (Eq. \ref{eq:de_alphas}) most of the parameter space does not allow for this cancellation.

Figure \ref{fig:triangle_all} also shows, in green, the constraints achievable after adding independent BAO and growth rate information from a DESI-like spectroscopic survey. As discussed in Section \ref{sssec:spec}, even though we observe a non-negligible improvement in the final uncertainties, the inclusion of spectroscopic information does not provide enough additional information to qualitatively change our results.

Finally, by comparing our forecast constraints with the analysis of \cite{Bellini:2015xja} with current data, we find that it will be possible to constrain $c_B$ and $c_M$ a factor of $\sim5$ better with next-generation surveys and, as reported above, $c_T$ will be measured with similar accuracy. It is worth noting that the specific constraints on $c_X$ depend mildly on the fiducial cosmology used in the forecast. The forecast uncertainties assuming the best-fit point of \cite{Bellini:2015xja} are shown in the bottom row of Table \ref{table:sigmas_exp}, and would correspond to a detection of deviations from $\Lambda$ GR with a significance above $7\sigma$.

\begin{table}
\centering
\begin{tabular}{c|c|c}
Parameter & Fixed params & 68\% uncertainty  \\
\hline
$c_B$ & $c_K,\,w$ & 0.128 \\
$c_B$ & $c_K$     & 0.161 \\
$c_B$ & None      & 0.167 \\
\hline
$c_M$ & $c_K,\,w$ & 0.065 \\
$c_M$ & $c_K$     & 0.070 \\
$c_M$ & None      & 0.072 \\
\hline
$c_T$ & $c_K,\,w$ & 0.146 \\
$c_T$ & $c_K$     & 0.151 \\
$c_T$ & None      & 0.151 \\
\hline
$c_K$ & $w$       & 3.11  \\
$c_K$ & None      & 3.13  \\
\hline
\end{tabular}
\caption{$1\sigma$ constraints for the Horndeski parameters for a combination of
         CMB-S4, LSST and SKA intensity mapping. Results are shown for $c_B$,
         $c_M$ and $c_T$, with $c_K$ and $w$ fixed or marginalized over. Results for
         $c_K$ are likewise shown for fixed and marginalized $w$.}
\label{table:sigmas_marg}
\end{table}

\subsection{Ultra-large scales and relativistic effects.}
It is also important to understand the dependence of the forecast constraints on the scales probed by each experiment. The interest is two-fold: firstly, a number of relativistic effects are known to leave an imprint on large-scale-structure observables on scales of the order of the horizon at the source redshift \cite{Alonso:2015sfa,Baker:2015bva,Renk:2016olm}, and their purely relativistic nature suggests that they may be relevant in constraining deviations from GR. Secondly, a large fraction of the sources of systematic uncertainty that Stage-IV surveys will be sensitive to, such as the problem of CMB and radio foregrounds, or star contamination in galaxy surveys, affect the measurement of density fluctuations on large angular and radial scales. This analysis therefore helps us quantify the amount of information lost by the anticipated loss of sensitivity on large scales.

\begin{figure}[t]
  \centering
  \includegraphics[width=0.48\textwidth]{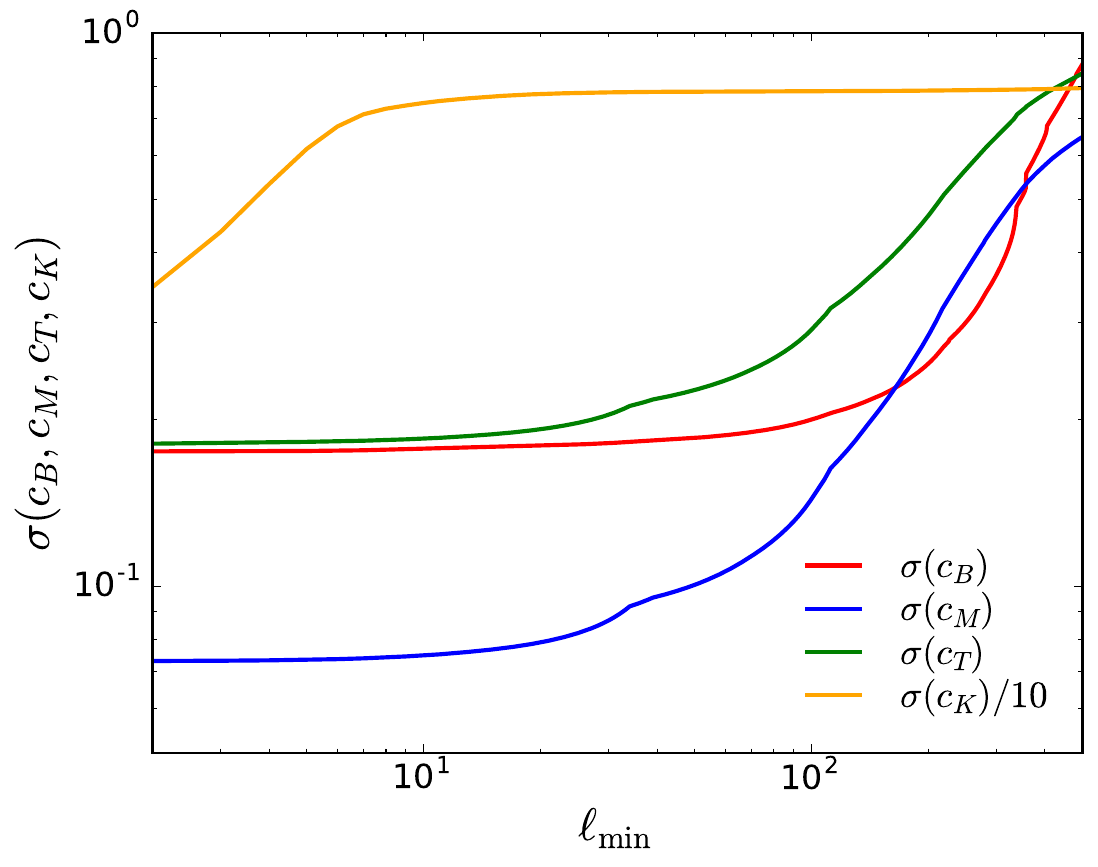}
  \caption{Constraints on the Horndeski parameter as a function of the minimum multipole $\ell_{\rm min}$ included in the analysis. Except for $c_K$, most of the information on the Horndeski parameters is encoded in the smaller scales, due to their higher statistical weight. Furthermore, we do not find a significant improvement in the constraints by including relativistic effects in galaxy clustering, given their relatively small contribution to the galaxy power spectrum and their redundance after including weak lensing observations.}
  \label{fig:lmin}
  \vspace{-1em}
\end{figure}

Figure \ref{fig:lmin} shows the $1\sigma$ uncertainties on the Horndeski parameters as a function of the minimum multipole (i.e. largest angular scale) included in the analysis. We see that, while most of the constraining power on $c_K$ come from multipoles $\ell\lesssim10$, it is the larger statistical power borne by small-scale perturbations (given the larger number of available small-scale modes) that drives the constraints on $c_B$, $c_M$ and $c_T$. As shown in Appendix \ref{app:quasi-static}, this is understandable, given that the leading order on sub-horizon scales, i.e.\ $ak/H\gg 1$, of the quasi-static parameters does not depend on $\alpha_K$.

It is possible to directly quantify the actual constraining power of the relativistic effects in large-scale structure \cite{Yoo:2009au,2011PhRvD..84d3516C,2011PhRvD..84f3505B}. In order to do so, we re-computed our forecasts after cancelling the contribution from these effects to the total fluctuations on number counts; specifically we discard contributions from the integrated Sachs-Wolfe, the Doppler correction to the Kaiser effect, the gravitational time delay, the effect of local gravitational potentials and the lensing magnification effect.  Comparing the resulting uncertainties with our nominal results we only observe a negligible improvement in the final constraints of up to $\sim1\%$, even in the case of $c_K$. Although this result may come as a surprise, given the direct relation of these effects with GR, it is actually understandable, given the small relative contribution of these effects to the overall fluctuations (e.g. see \cite{Baker:2015bva}), which is 
comparable to the effect of primordial non-Gaussianity on the large-scale clustering of a biased tracer with bias $b\sim2$ and $f_{\rm NL}\sim1$.

Unlike other relativistic effects, the impact of lensing magnification correction on galaxy number counts persists on small scales. In Horndeski models it leads to distinct signatures on the cross correlation between galaxies with large redshift separations \cite{Renk:2016olm}. Nonetheless, other probes of the gravitational potential such as shear measurements or CMB lensing probe the lensing potential with much higher signal-to-noise ratio. Hence, including or removing the lensing effect does not improve the sensitivity when clustering measurements are combined with lensing. For an LSST clustering-only forecast the constraints on the Horndeski parameters benefit from the inclusion of lensing magnification, with the 1-sigma bounds on $c_B,\,c_M,\,c_T$ improving respectively by a factor 7, 2 and 3, but still considerably worse than the full LSST results including shear measurements.

\subsection{Small Scales}\label{ssec:smallscales}
\begin{figure}[t]
      \centering
      \includegraphics[width=0.48\textwidth]{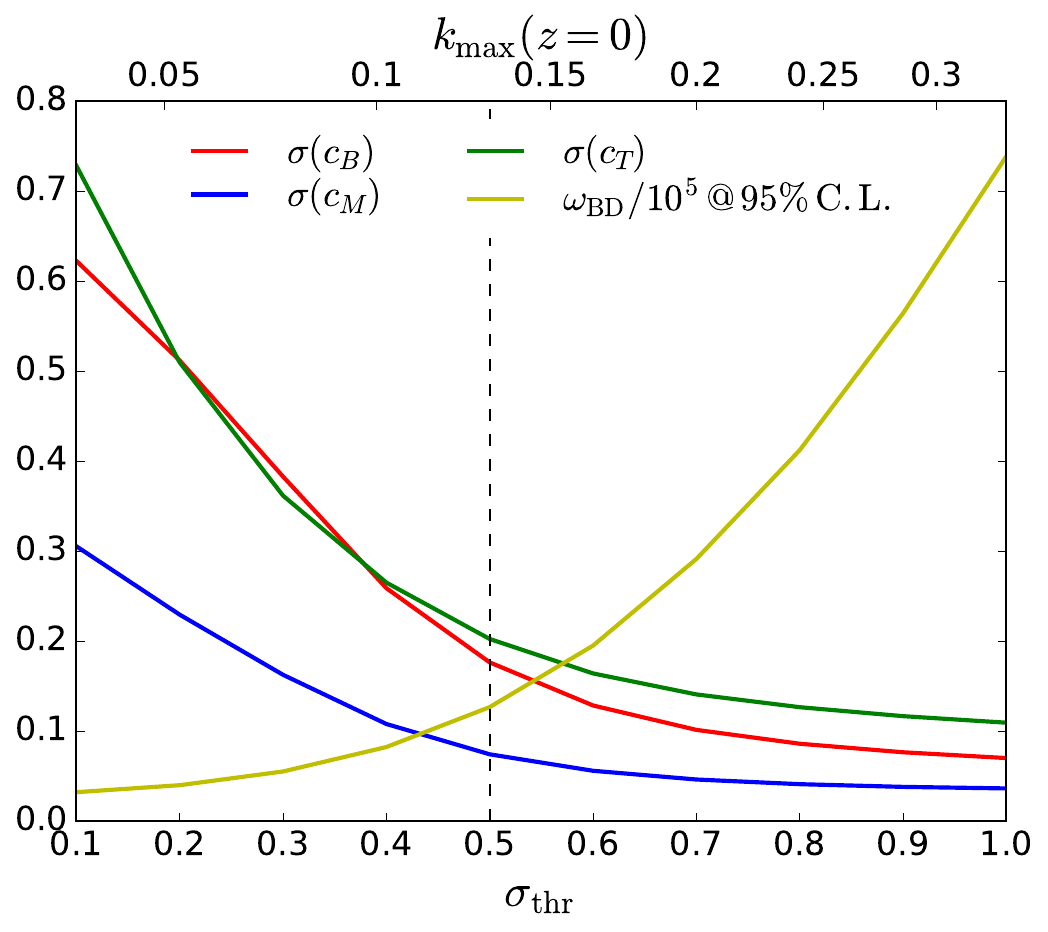}
      \caption{Forecast constraints  on $c_M$, $c_B$, $c_T$ and $\omega_{\rm BD}$ (the latter given as the upper bound at 95\% C.L.) as a function of the maximum mode included, $\ell_{\rm max}$, set by maximum amplitude of clustering $\sigma_{\rm thr}$ as described in Section \ref{sec:surveys}.}
      \label{fig:nl}
      \vspace{-1em}
    \end{figure}
\begin{figure}[t]
      \centering
      \includegraphics[width=0.48\textwidth]{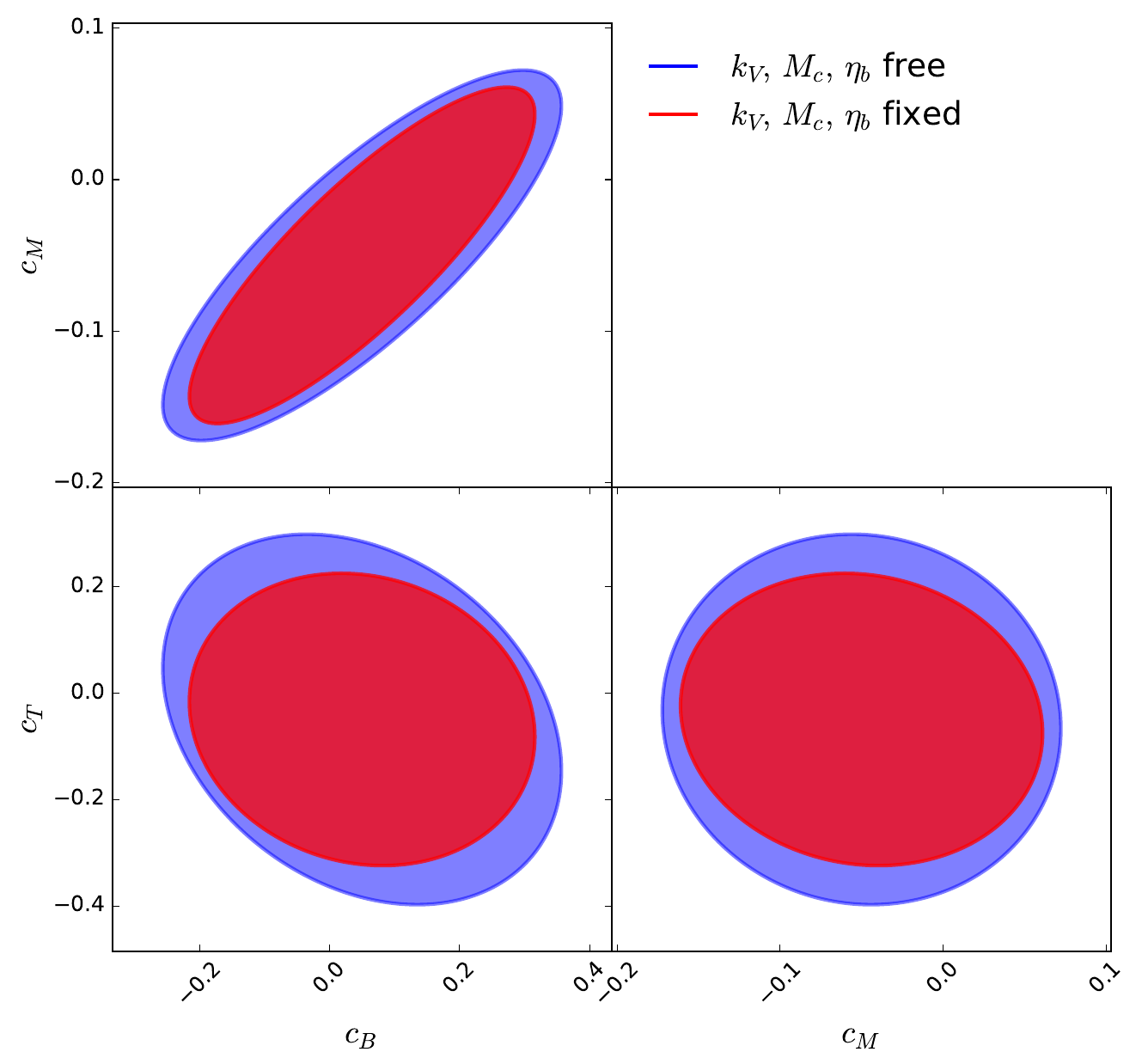}
      \caption{Forecast constraints  on $c_M$, $c_B$ and $c_T$ assuming perfect knowledge of baryonic physics and screening effects (red ellipses) and marginalising over them (blue ellipses) as parametrised in Section \ref{sec:horndeski}.}
      \label{fig:kvbar}
      \vspace{-1em}
    \end{figure}
Smaller scales have smaller cosmic variance and hence much larger statistical weight than large scales. However it is on small scales that non-linear gravity and baryonic effects will play a significant role. For example, at $z=0$, non-linear corrections to the power spectrum are of order $1\%$ at $k\sim 0.01$ \cite{Baldauf:2016sjb}. There is concerted effort under way to accurately model non-linear corrections to sufficient precision so that one will be able to use small scale modes in future analysis (attempts at doing this with current data can be found in \cite{Reid:2014iaa} and a variety of approaches can be found in \cite{Crocce:2005xy,Carrasco:2012cv,Blas:2015qsi}). There have also been attempts at constructing phenomenological models which would include baryonic effects, including feedback \cite{2015MNRAS.450.1212H,Mead:2015yca}.

While there is some hope that, within the context of standard models (with a cosmological constant and evolving under GR), it should be possible to harness some non-linear modes, the case for models with modified gravity is less promising. A suite of N-body codes exists for particular subclasses of theories beyond general relativity, and there has been an attempt at cross-calibrating them to better than $1\%$ on a wide range of scales \cite{Winther:2015wla}. Yet, the range of models which have been compared is still relatively restricted and is not sufficiently general that we have a firm understanding of the non-linear regime. 

\begin{figure*}
      \centering
      \includegraphics[width=0.7\textwidth]{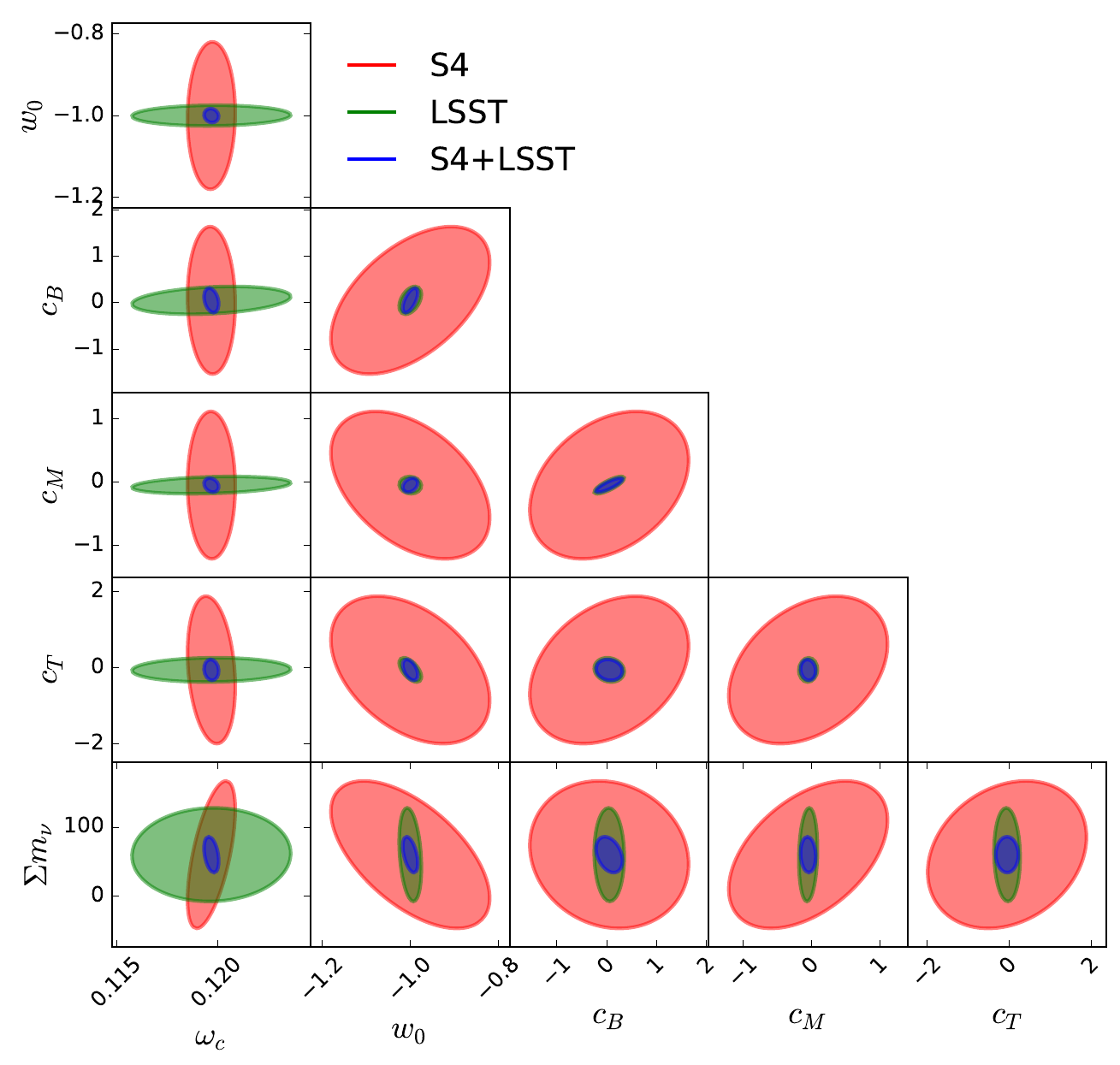}
      \caption{1$\sigma$ cosmological constraints achievable by a Stage-4 CMB experiments (red), LSST (green) and the combination of the two (blue).}
      \label{fig:S4LSST}
      \vspace{-1em}
    \end{figure*}

A further complication of these models is the accurate modeling of non-linear screening mechanisms. Lacking an analytical treatment, we have opted modeled these effects as an scale dependent modulation of the ``$\alpha$''-functions, as described in equation \ref{eq:vainshtein_cutoff}, to effectively suppress any deviation from GR in the non-linear regime while retaining the information gain of non-linear scales on the remaining cosmological parameters. Note that including non-linear scales without accounting for screening could severely overestimate the constraining power of future surveys. Note that the inclusion of a finite cutoff $k_V$ eliminates the effects of modified gravity on smaller scales, as expected due to screening mechanisms, but also introduces a feature in the matter power spectrum that can lead to a stronger measurement of the gravitational parameters. Although the signal-to-noise ratio is higher, this procedure actually slightly improves the constraints on the Horndeski parameters. The reason is that the screening cutoff effectively introduces a feature in the power spectrum that becomes sharper with stronger modifications of gravity. Similarly to the BAO feature, this feature is not degenerate with galaxy bias and its detection is more robust than the broadband amplitude. The fact that the constraints improve indicates that the existence of a screening feature  effectively compensates the loss of signal-to-noise from weakening the modifications on smaller scales.

To explore the dependence of the constraints on the choice of smallest scale to include in the analysis, we have recomputed our forecasts for more optimistic and pessimistic values of $\sigma_{\rm thr}$ (see Eq. \ref{eq:sigma_thr}). Figure \ref{fig:nl} shows the dependence of the final constraints on this choice. Even though the final uncertainties on e.g.\ $c_B$ or $c_T$ may decrease by a factor of $\sim2$ with respect to the constraints found for our fiducial choice (displayed by a vertical dashed line), we find that significant information can still be gained assuming a correct modelling of the matter power spectrum down to scales $k\sim0.1\,h{\rm Mpc}^-1$ at $z=0$. 

We have also quantified the impact of uncertainties on parametrization of the non-linear screening and the effect of baryonic physics on the power spectrum by marginalising over the values of $M_c$, $\eta_b$ (see \cite{2015JCAP...12..049S}) and $k_V$. Figure \ref{fig:kvbar} shows the 1$\sigma$ constraints on the Horndeski parameters assuming perfect knowledge of these parameters (red ellipses) and marginalising over them (blue ellipses). The degradation in the final constraints is kept at a reasonable level due to the conservative cut on small scales used in this work.

\subsection{The relative importance of the different surveys.}
One can broadly characterise a cosmological survey in terms of its ability to measure two main observables: the  angular diameter distance relation and growth of structure. The former will be sensitive to the background expansion and therefore, the equation of state $w$, while the latter will depend on all parameters. It should also be possible to constrain the shape of the power spectrum as a function of redshift and thus pick out scale-dependent effects on the growth rate (e.g. the braiding scale in Eq.\ \ref{eq:k_braiding} and the $k$-dependence in the quasi-static parameters, Appendix \ref{app:quasi-static}). But this will be, for now, a subdominant effect and, in the case of a galaxy redshift survey, is very sensitive to assumptions about bias.

At early times, a measurement of the primary anisotropies in the CMB will serve as an anchor for both distance measurements as well as for the growth rate: it fixes the angular diameter distance at $z\simeq 1000$ as well as the overall primordial amplitude of fluctations. Per se it will not tightly constrain the gravitational parameters but it will play a crucial role in breaking degeneracies. In addition, S4 will supply us with a high-significance map of the projected matter fluctuations (with a radial kernel peaking around $z\sim 2$) via weak lensing. As such it will help to calibrate measurements of the growth rate at lower redshift as well as to pin down the neutrino mass.

Complementing early-time constraints from the CMB are late-time measurements of large scale structure from galaxy clustering and weak lensing with LSST as well as intensity mapping with SKA. Specifically, an intensity mapping survey such as SKA1-IM will give us a biased measurement of the matter power spectrum as a function of redshift and therefore a distance measurement via the baryon acoustic oscillation features. To a lesser degree of importance, it will also give us a measurement of the growth rate via redshift-space distortion, but only if an independent measurement of the background HI temperature can be made. The LSST survey will have two complementary data sets. On the one hand it will supply us with a map of the galaxy distribution and therefore a measurement of the angular diameter distance as well as a low-significance measurement of redshift-space distortions over a range of redshifts. On the other hand it will supply us with a tomographic set of weak lensing maps which will give us an unbiased 
measurement of the growth of structure through the matter power spectrum.

In Figure \ref{fig:S4LSST} we can see the important role that the complementarity between early-time and late-time constraints. It is already well established that the CMB and late time measurements (such as LSST) will combine to supply powerful constraints on the equation of state, the matter density and neutrino masses. The degeneracies broken by this combination thus improve the uncertainties on the gravitational parameters, as can be seen in the $\sim30\%$ increase in the figure-of-merit shown in the last column of Table \ref{table:sigmas_exp} between rows 2 an 4. There are two reasons for this. First of all, the CMB will anchor the distance measurement so that with LSST, it is possible to greatly reduce the uncertainty on $w$. This will feed into the degeneracy between $c_M$, $c_B$ and $w$ in the growth rate. Second, the CMB weak lensing will pin down the matter power spectrum, complementing the measurements of the growth rate via the LSST clustering survey and the the matter power spectrum via the LSST 
weak lensing survey.

\begin{figure}[t]
      \centering
      \includegraphics[width=0.48\textwidth]{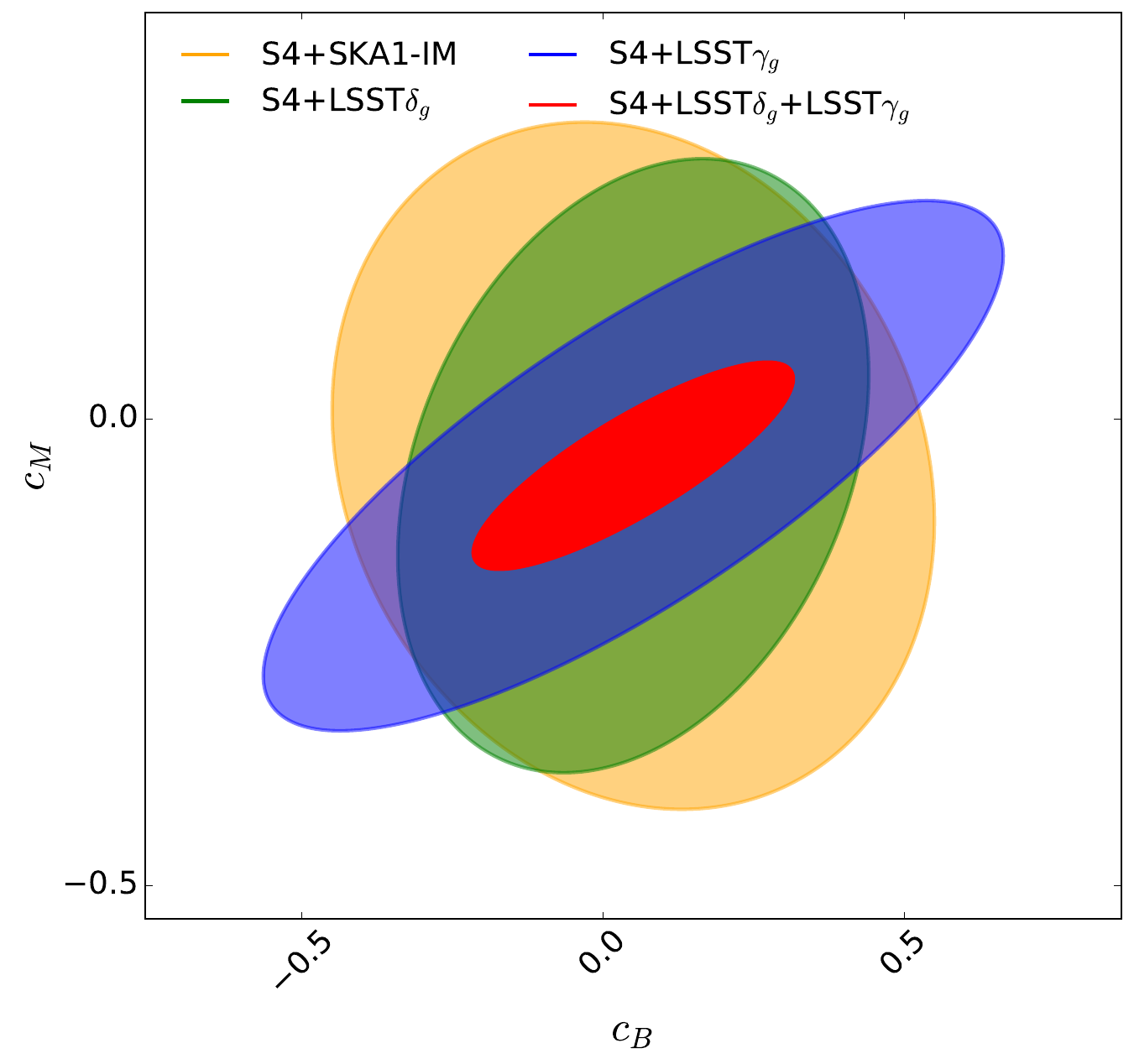}
      \caption{Complementarity between clustering observables (SKA intensity mapping in orange and LSST clustering in green), providing measurements of expansion like the BAO scale, and weak lensing (LSST shear in blue), providing measurements of the growth of structure. The final constraints improve massively after combining both observables (red).}
      \label{fig:LSSTdg}
      \vspace{-1em}
    \end{figure}
As independent probes of geometry (through the BAO scale) and growth (through the power spectrum), it is also useful to look at the individual contributions of galaxy clustering and cosmic shear to the constraints.
In Figure \ref{fig:LSSTdg} we show how clustering and shear complement each other in the $c_B$-$c_M$ plane. In particular, the ability of the shear catalogue to give us an unbiased measurement of the matter power spectrum as a function of time leads to an improved measurement of the standard cosmological parameters and the dark energy equation of state, and thus helps break some of the degeneracies that arise from galaxy clustering alone. The figure also shows how, unsurprisingly, the contributions the two clustering probes individually (intensity mapping and LSST clustering) are similar.

\subsection{More complex time dependence.}\label{ssec:timedep}
\begin{table}
\centering{
  \renewcommand*{\arraystretch}{1.6}
\begin{tabular}{|C{1cm}|C{1.4cm}|C{1.2cm} | C{1.4Cm}|C{1.2cm}|}
\hline
\multirow{2}{*}{Term}
           & \multicolumn{2}{c|}{$\propto\Omega_{\rm DE}$}  & \multicolumn{2}{c|}{$\propto{\rm tanh}$} \\
\cline{2-5}
           & $\sigma_b$ (early)  & $\sigma_c$ (late)  & $\sigma_b$  (early)  & $\sigma_c$ (late) \\
\hline 
$\alpha_K$  & 0.29              & 3.5  & 0.15               & 1.6       \\
$\alpha_B$  & $1.3\times10^{-3}$& 0.17 & $2.3\times10^{-3}$ & 0.21      \\
$\alpha_M$  & $3.6\times10^{-5}$& 0.072 & $3.6\times10^{-5}$ & 0.013      \\
$\alpha_T$  & 0.12              & 0.18  & $2.7\times10^{-2}$ & 0.070      \\
\hline
\end{tabular}}
\caption{$1\sigma$ constraints on the Horndeski parameters $c_X$ (late-time) and $b_X$ (early-time). Columns 2 and 3 give the errors 
         for the parameterization proportional to the normalized dark energy density, Eq. \ref{eq:param_omega}. Columns 4 and 5 are for the time parametrization given by Eq. \ref{eq:timedep}, described in Section \ref{ssec:timedep}. In the latter case, forecasts include marginalization over the time-dependence parameters $z_H$ and $\Delta z_H$ (see Eq. \ref{eq:timedep}).}
\label{table:sigmas_td}
\end{table}

Throughout our analysis we have assumed a simplified fiducial model in which the time evolution of the $\alpha$s is tightly correlated with the emergence of an accelerated expansion phase. Given that our goal is to forecast how our constraints on the $\alpha$s will improve with Stage IV and given that current constraints are undertaken with such an assumption, we believe this is a sensible approach. The forecast constraints we obtain are meaningful and give us an idea of how much better constraints on Horndeski theory will be.

Nevertheless, we are well aware that, when focusing on specific models, our fiducial parametrization may not be ideal. It does correctly capture the time evolution for a large subclass of Horndeski models but it has been shown that for some subsets of model space, it is a poor approximation \cite{Linder:2015rcz}. In fact, as we will see, in the case of the simplest non-trivial Horndeski model - Jordan-Brans-Dicke theory - the $\alpha$s are approximately constant and, thus, their evolution is completely decoupled from the onset of accelerated expansion. Moreover, in models based on a covariant Lagrangian, such as Equation \ref{eq:L_horndeski}, the background and perturbation evolution will depend on the same set of parameters, and their evolution will be related. In contrast, we are effectively decoupling the background ($w(z)$) from the growth ($\alpha_X(z)$), effectively granting the model more freedom to simultaneously fit different observables.

A favoured approach to achieve full generality is to construct an orthogonal basis which is completely agnostic about the time evolution of the $\alpha$s using, for example, Principal Component Analysis or some form of eigenbasis \cite{Pogosian:2010tj,Hojjati:2011xd,Abazajian:2016yjj}. This approach is systematic and, in some sense, complete, and will clearly be useful when actually analyzing the data. However it is very difficult to interpret forecasted errors on such parametrizations in a meaningful way and so, in what follows, we will choose an alternative parametrization which incorporates some features that escape our fiducial model. Doing so will allow us to assess how dependent our forecasts are on our assumptions.

A first, straightforward modification is to incorporate a constant term that allows us to forecast constraints from epochs prior to the accelerated expansion era. This allows us to generalize the idea of early dark energy \cite{Pettorino:2013ia} and to place bounds on the properties of gravity at early times. Models that attempt to unify cosmic acceleration in inflation and the late Universe will be severely constrained by the data if they present with residual departures from GR between those stages. Thus, we start by reintroducing the constant term in our parametrization, given by $b_X$ in Eq. \ref{eq:param_omega}. This term modifies the dynamics throughout cosmic history (see \cite{Lima:2016npg,Perenon:2016blf} for the effect of modifications only on the matter era and \cite{Brax:2013fda} for CMB effects in a reduced parameterization). For small early-time departures from GR the super-horizon adiabatic perturbations evolve as in the standard case, allowing us to set the standard initial conditions \cite{
earlyMG:2016}. We shall see, when assessing constraints on Jordan-Brans-Dicke theories, that indeed these early-time, non-trivial values of $\alpha$ are present.

\begin{figure}[t]
      \centering
      \includegraphics[width=0.48\textwidth]{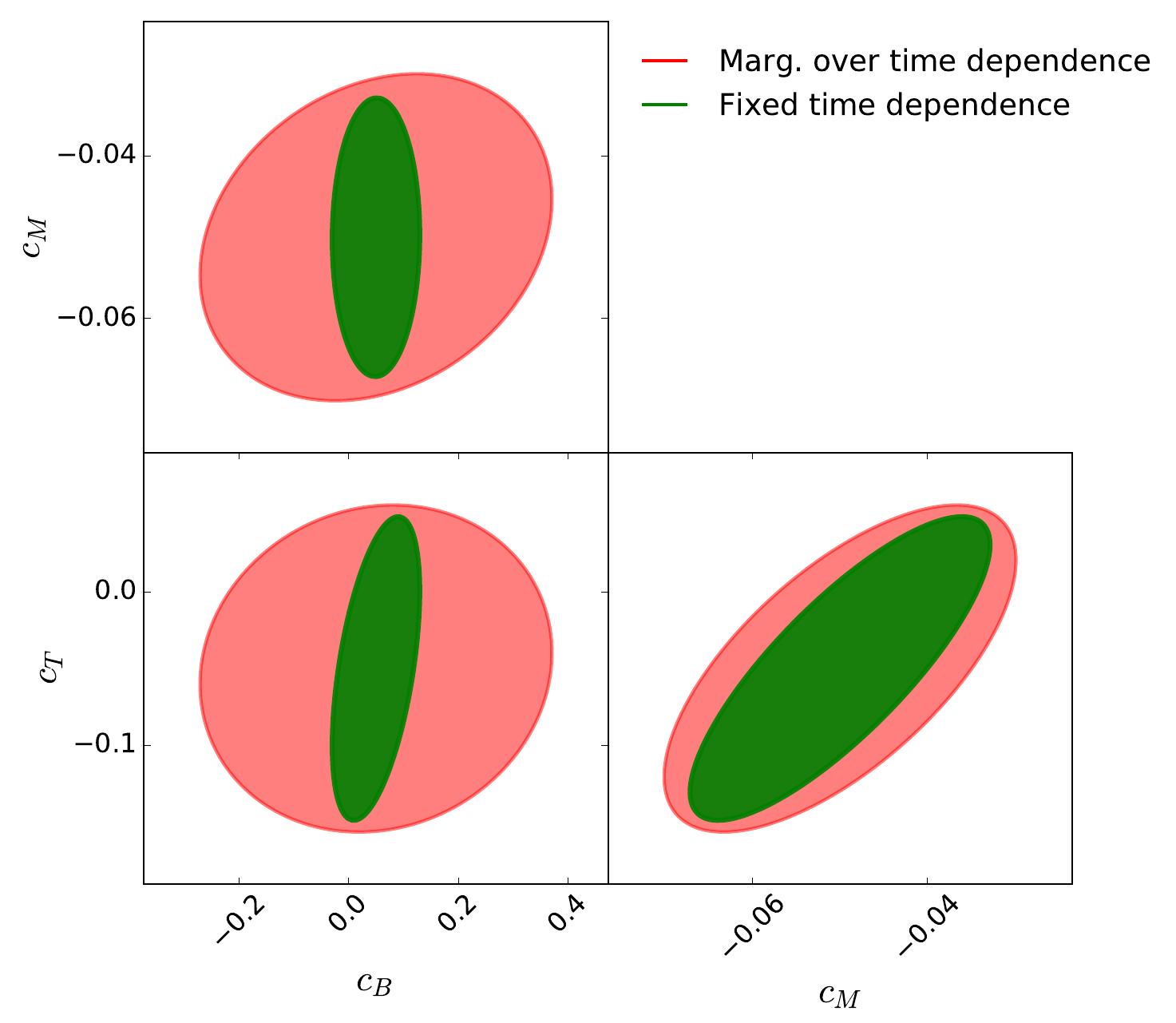}
      \caption{68\% constraints for the Horndeski parameters for a fixed time dependence (green) and marginalized over the time-dependence parameters $z_H$ and $\Delta z_H$ (see Eq. \ref{eq:timedep}).}
      \label{fig:zdep}
      \vspace{-1em}
    \end{figure}

Additionally, we have also considered an alternative parametrization where the transition between the early-time and late-time terms is characterized by a redshift $z_H$ and a transition width $\Delta z_H$:
\begin{equation}\label{eq:timedep}
 \alpha_X=b_X+(c_X-b_X)\frac{1-{\rm tanh}[(z-z_H)/(2\Delta z_H)]}{1+{\rm tanh}[z_H/(2\Delta z_H)]}.
\end{equation}
By marginalizing over $z_H$ and $\Delta z_H$, this model thus allows us to assess the impact of uncertainties on the precise time evolution of the Horndeski parameters on the final constraints on their early-time and late-time values. Here we will use fiducial values for these parameters $z_H=0.5$, $\Delta z_H=0.5$. The corresponding time evolution of the $\alpha$s roughly mimics that of $\propto\Omega_{\rm DE}$, although the effects of the late-time modifications are more prominent for a slightly larger period of time.

Columns 2 and 4 in Table \ref{table:sigmas_td} show the forecast 1$\sigma$ uncertainties from the combination of S4 and LSST for the early-time parameters for both parametrizations, while the late-time constraints are shown in columns 3 and 5. We find that early-time modification are much more tightly constrained than late-time modifications. This is to be expected, as early modifications will have an impact on CMB observables and induce modification to the growth rate which will then persist over a much longer period of time. Furthermore, the effects of late and early times are almost entirely uncorrelated. This can be seen if we compare the fourth row of Table \ref{table:sigmas_exp} with the fourth column of Table \ref{table:sigmas_td}, where the uncertainty on $c_X$ account for marginalization over $b_X$. 

It is also worth noting that, while the achievable uncertainties on both the early and late-time parameters are similar in both models, the constraints improve for the new parametrization (see below for the particular case of $c_B$). The reason for this is twofold: first, the effect of the late-time terms is present for a larger period of time in this model. Second, the time derivative of the $\alpha$s may be more pronounced in this model (depending on the choice of $\Delta z_H$), and since this time derivative enters in the evolution equations the constraints are tighter.

Finally, Figure \ref{fig:zdep} shows the effects of marginalising over the time-evolution parameters $z_H$ and $\Delta z_H$ (without any prior) on the final constraints on the late-time parameters. While the effect on $c_M$ and $c_T$ is small, the uncertainty on $c_B$ increases by a factor $\sim3$. This is likely because the perturbation equations depend on the time derivative of $\alpha_B$ (as well as $\alpha_K$) and hence changing the width of the transition is more degenerate with $c_B$ than $\alpha_M,\alpha_T$, whose time derivative does not enter the perturbation equations (cf. appendix A2 in Ref. \cite{Zumalacarregui:2016pph}). This shows that, when constraining general classes of models, it will be important to account for possible uncertainties in the time evolution of the modifications, although tight constraints can still be achieved after this marginalization.

\subsection{Constraints on the Jordan-Brans Dicke Theory}\label{ssec:bd}
\begin{figure}[t]
      \centering
      \includegraphics[width=0.48\textwidth]{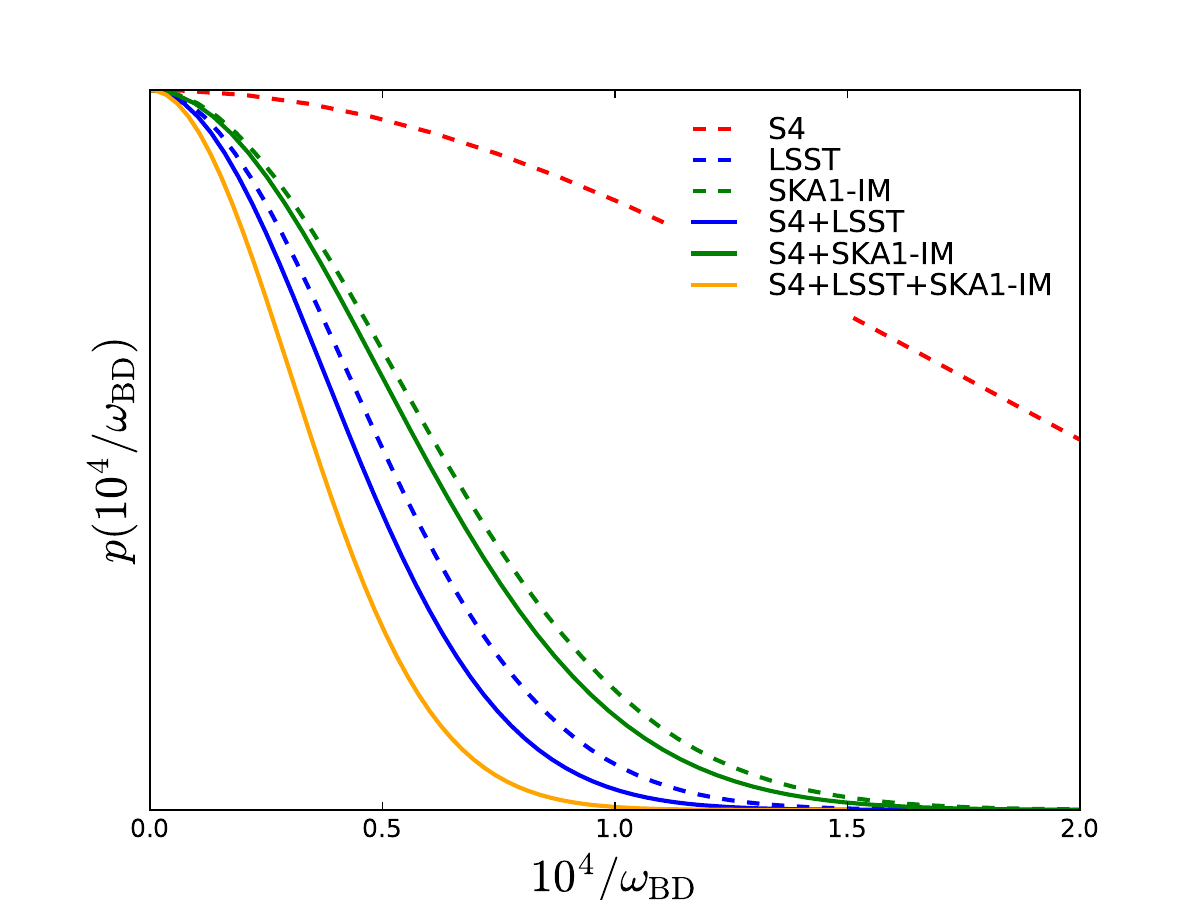}
      \caption{Marginalized distribution for the inverse of the Brans-Dicke parameter $\omega_{\rm BD}$ for different combinations of experiments. The constraints are mostly driven by late-time probes.}
      \label{fig:obd}
      \vspace{-1em}
    \end{figure}
We now focus on a very specific class of models: Jordan-Brans Dicke theory \cite{Brans:1961sx} . One can think of this as a very restricted set of priors on the $\alpha$s in the Horndeski theory. In fact we can express the $\alpha$s in terms of the background scalar field evolution, $\phi(t)$ and the Brans-Dicke parameter, $\omega_{\rm BD}$ as
\begin{eqnarray}
\alpha_M&=&\frac{d\ln \phi}{d\ln a}, \nonumber \\
\alpha_B&=&-\alpha_M, \nonumber \\
\alpha_K&=&\omega_{\rm BD}\alpha^2_M, \nonumber \\
\alpha_T&=&=0.
\end{eqnarray}
In principle it should be possible to further restrict the dependence of the $\alpha$s on $\phi$ by using the fact that $\phi(t)$, in certain situations, will lock onto a tracking solution. In an Einstein-De-Sitter universe we have $\phi=\phi_0a^{1/(\omega_{\rm BD}+1)}$ \cite{Nariai:1969vh} where $\phi_0=(2\omega_{\rm BD}+4)/(2\omega_{\rm BD}+3)$ and therefore $\alpha_M=1/(\omega_{\rm BD}+1)$. In other words, the  Jordan-Brans Dicke theory corresponds to a point in the space defined by the $\alpha$s.

In the presence of dark energy, characterized by an equation of state $w$, there will be very slight deviations from the attractor due to late-time acceleration of the scalar field \cite{Lima:2015xia}, although it may still be possible to construct an accurate, perturbative solution to the time evolution of $\phi$ that only depends on $\omega_{\rm BD}$ and $w$. If that is the case, them the $\alpha$ will be uniquely determined by $\omega_{\rm BD}$ and $w$. Note that, for simplicity, we will set $w=-1$ in this section. 

We carry out a Fisher matrix analysis using $1/\omega_{\rm BD}$ as a parameter which we vary around the fiducial point $\omega_{\rm BD}=10^5$ (we verified that the forecasts do not depend on this choice). Assuming that the uncertainty on $1/\omega_{\rm BD}$ is Gaussianly distributed, it is straightforward to translate the 1$\sigma$ uncertainty on $1/\omega_{\rm BD}$ into a 95\% C.L. lower bound for $\omega_{\rm BD}$. The second column in Table \ref{table:sigmas_exp} shows the forecast values of this lower bound for a variety of combinations of data sets, and Figure \ref{fig:obd} displays the forecast 1D distribution of $1/\omega_{\rm BD}$ for a similar range of observations. For a start we can see that the combined, most optimistic constraint is remarkable: we will be able to place a lower bound, $\omega_{\rm BD}>1.7\times10^{4}$, which is comparable to current millisecond pulsar constraints and just under current constraints from Solar system. 

The fact that different data sets probe very different regimes plays a crucial role in our findings. Unlike in the case of our simplified parametrization of the $\alpha$s (in terms of $c_M$, etc), in this case, for a given value of $\omega_{\rm BD}$, the $\alpha$s are non zero throughout cosmic history. We see then that combining the CMB S4 data, which primarily constrains the $\alpha$s at $z\sim1000$ with late-time constraints from LSST galaxy and weak lensing surveys will improve the constraint on $\omega_{\rm BD}$ by a factor of 4 or more. It is also interesting to note that, independently, both SKA-IM and LSST give us a bound on $\omega_{\rm BD}$ of order $10^4$. Given that each of these data sets will be affected by their own set of systematics, this will allow us a stringent cross-check on the bounds. 

Finally, it is important to check how our ability to model non-linear scales will affect our constraint. In Figure
\ref{fig:nl} we show the projected lower bound on $\omega_{\rm BD}$ as function of the maximum wavenumber
(or density threshold). Reducing what corresponds to $k_{\rm max}$ at $z=0$ from $0.1$ Mpc$^{-1}$ (our fiducial value) to $0.05$ Mpc$^{-1}$, degrades the constrain on $\omega_{\rm BD}$ by more than a factor of $3$. If cosmological constraints are to be competitive with other, more local, constraints, it will be important to model non-linearities accurately for Jordan-Brans-Dicke theory; given its dependence on one (constant) parameter this should be a much simpler problem than in the general Horndeski scenario. Moreover, this  theory does not have a screening mechanism and hence reliably modeling non-linear scales significantly improves the constraints, as can be seen in Fig. \ref{fig:nl}.

\section{Discussion}\label{sec:discussion}

In this paper we have looked forward at what might be achievable with future surveys, focusing on constraints on scalar-tensor theories described by a broad subclass of Horndeski theories as well as a very specific model: Jordan-Bran-Dicke theory. Our results have been enlightening and we can summarise them as follows:
\begin{itemize}
\item Constraints on the Horndeski parameters will greatly improve. But, given the weak effect $\alpha_K$ has on the data, we only expect precise constraints on $\alpha_M$, $\alpha_B$ and $\alpha_T$.
\item We find that, in the best case scenario, through a combination of Stage-IV CMB, photometric and spectroscopic surveys, constraints on $\alpha_M$ and $\alpha_B$ will improve (relative to current constraints) by a factor of $5$ to $\sigma(c_M)\simeq 0.06$ and $\sigma(c_B)\simeq0.12$. Furthermore, $\alpha_T$ will join the ranks of the well-constrained parameters, with $\sigma(c_T)\simeq 0.15$.
\item Relativistic effects and, more generally, ultra-large-scale modes, will have negligible statistical weight on the constraints on $\alpha$s compared to smaller-scale fluctuations. 
\item There are correlations between the gravitational parameters, $\alpha_{B,M,T}$ and the more conventional cosmological parameters, $w$ and $\omega_c$, which can be understood in terms of their joint effects on the growth of structure.
\item There is no apparent degeneracy between modified gravity and the sum of neutrino masses.
\item The combined Stage-IV surveys will lead to a lower bound on the Brans-Dicke parameter $\omega_{\rm BD}>1.7\times 10^4$, which is comparable with current millisecond pulsar constraints and of the same order of current Solar System constraints.
\item It will be important to find an accurate method for modelling non-linearities, although restricting the analysis to purely linear modes still leads to moderately tight constraints. Focusing on the Jordan-Brans-Dicke theory, we find that the degradation in the lower bound of $\omega_{\rm BD}$ can be substantial, even if we discard a moderate number of non-linear modes.
\item The model for the time evolution of the $\alpha$s can have a substantial impact in the final constraints. While the fiducial model we use here captures the correct behaviour at low redshift for a large sub-class of the Horndeski theories, the accuracy of the forecasts and future constraints would be greatly improved with
a parametrization which reflects the correct, theoretically sound, underlying behaviour of the parameters. 
\end{itemize}

There are a few comments we can make, building on our conclusions. For a start, the fact that ultra-large scales play a negligible role in constraining the gravitational parameters lends credence to the approach of \cite{Pogosian:2016pwr} and \cite{Perenon:2016blf}. There, the authors suggest that, by looking at the values of the quasi-static parameters, it is possible to efficiently rule out large swathes of the Horndeski family. In a sense, by establishing a clear classification of the effects of Horndeski theories on the effective Newton's constant and the gravitational slip, it is possible to make definitive quantitative statements about the values and ranges of the $\alpha$s which are observationally viable.

It is interesting to compare different combination of surveys and their ability to constrain the gravitational parameters. We can define a figure of merit (FoM) from the Fisher matrix%
\footnote{Our definition ensures that the FoM scales as a surface, rather than the multi-dimensional volume of the parameter space (cf. \cite{Nesseris:2012cq,Alarcon:2016bkr}). This facilitates comparion of FoMs involving different number of parameters.}
\begin{equation}
 {\rm FoM}_S = \left[\det\left(F^{-1}\right)_S\right]^{-1/\dim(S)},
\end{equation}
Where $S$ indicates the subspace of parameters that enter the figure of merit (all others marginalized), which we have chosen to be the set of currently constrainable Horndeski parameters $\{c_B,c_M,c_T\}$ \cite{Bellini:2015xja}.
We do not include the expansion rate of the Universe (captured by the commonly used $w-w_a$ FoM) to focus instead on the impact of gravity on LSS.
The values for this figure of merit forecast for the different experiments discussed in this work are shown in the last column of Table \ref{table:sigmas_exp}. It clearly displays the benefit of combining different probes to test gravity: the figure of merit for S4+LSST+SKA1-IM increases by two orders of magnitude relative to single-experiment S4 and SKA1-IM, and roughly doubles with respect to LSST-only. Further addition of a DESI-like spectroscopic survey increases this FoM by about 15\%. These values show a very significant improvement with respect to current constraints. By restricting ourselves to the $c_B-c_M$ plane, constrained by \cite{Bellini:2015xja}, we observe an increase of a factor $\sim40$ in the figure of merit between current constraints and those achievable by the combination of our three main experiments.

We have restricted ourselves to minimally coupled, strictly Horndeski terms. It is straightforward to enlarge our parameter space, as has been done in \cite{Gleyzes:2015rua} and \cite{D'Amico:2016ltd}. In fact, it is now possible to go beyond scalar-tensor theories and consider vector-tensor (such Einstein-Aether or Maxwell-Proca) or tensor-tensor (such as bigravity models) self-consistently on linear scales \cite{Lagos:2016wyv,Lagos:2016gep}. Furthermore, we haven't included spectroscopic redshift surveys in a completely consistent way. As explained above, a completely consistent yet computational tractable way of including future redshift surveys is still lacking although, with inexorable improvement of computational speed and memory, may be achievable within the next few years. However, the result that ultra-large scales do not play a significant role in the constraints means that it should be possible to construct an approximate, but far more efficient, forecasting apparatus using three dimensional power spectra (as opposed to angular cross-power spectra).

Our results are also incomplete with regards to the treatment of systematic uncertainties. Even though, as stated in the text, the availability of cross-correlations between different surveys and probes ensures a robust self-calibration of sources of systematic uncertainty such as the effect of intrinsic alignments, a full treatment of shape-measurement systematics in weak lensing, or a complete characterisation of photo-$z$ uncertainties would allow us to place forecasts on scalar-tensor theories at the same level as the work currently done for standard cosmological scenarios. We leave this for future work, but expect these effects to somewhat weaken our forecasts.

On other hand, even though we have considered most of the relevant overlapping Stage-IV surveys, thus exploiting the complementary coverage of the same patch of the sky by multiple tracers of the matter density field, there are other cosmological probes that might be used to further constrain gravity. Notable examples are peculiar velocities (nearby, via Tully-Fisher measurements or distant, via the kinetic-Sunyaev-Zeldovich effect), cluster number counts (using optical, X-ray or Sunyaev-Zeldovich measurements) and distance measurements from supernovae and strong lenses. All of these will add further statistical weight to our forecast constraints and could reduce the uncertainties on the gravitational parameters.

What is clear is that the future of gravitational physics in the context of observational cosmology is very promising. Over the next decade, constraints on cosmological scales will match in precision those on astrophysical scales. This will place GR as one of the most thoroughly tested theories in the physical canon.

\section*{Acknowledgments}
We acknowledge discussions with T. Baker, E. Calabrese, M. Lagos, N. Lima, L. Lombriser, S. Nesseris, J. Noller and A. Silvestri. We are grateful to A. Avilez and C. Skordis for help in testing our implementation of Brans-Dicke models. DA is supported by the ERC, STFC and BIPAC. PGF acknowledges support from STFC, BIPAC, the Higgs Centre at the University of Edinburgh and ERC. EB is supported by the ERC and BIPAC.
 
\bibliography{paper}

\appendix

\section{The Quasi-Static Limit}
  \label{app:quasi-static}
 Consider the quasi-static parameters introduced in Eq.~\ref{eq:qs-parameters}
 \[
\begin{cases}
\gamma & =\,\gamma_{0}\left(t\right)+\gamma_{1}\left(t\right)\left(\frac{\cal H}{k}\right)^{2}\\
G_{eff} & =\,G_{0}\left(t\right)+G_{1}\left(t\right)\left(\frac{\cal H}{k}\right)^{2}.
\end{cases}
\]
We can relate them to the Horndeski parameters as follows.
 At leading order we have
\begin{align}\label{eq:leading_QS}
\gamma_{0}\equiv\, & \frac{\beta_{3}}{\beta_{1}+\beta_{2}},\\
G_{0}\equiv\, & \frac{2\left(\beta_{1}+\beta_{2}\right)}{M_{*}^{2}\left[2\beta_{1}+\left(2-\alpha_{\textrm{B}}\right)\beta_{2}\right]}.
\end{align}
At second order in $\left({\cal H}/k\right)$ we have
\begin{align}
\frac{\gamma_{\text{N1}}}{\alpha_{\textrm{M}}-\alpha_{\textrm{T}}}\equiv\, & 3\alpha_{\textrm{B}}\beta_{2}\left[1+\frac{H^{\prime\prime}}{aHH^{\prime}}\right]\frac{H^{\prime}}{aH^{2}} \\ &-3\beta_{1}\left[\beta_{3}-2\left(\alpha_{\textrm{M}}-\alpha_{\textrm{T}}\right)\frac{H^{\prime}}{aH^{2}}\right]\\
 & -\alpha_{\textrm{K}}\frac{\beta_{3}}{aH}\left[\frac{H^{\prime}}{H}+\frac{\alpha_{\textrm{K}}^{\prime}}{\alpha_{\textrm{K}}}\right] \\ &-\left(3\alpha_{\textrm{B}}+\alpha_{\textrm{K}}\right)\beta_{3}\left[3+\alpha_{\textrm{M}}+\frac{H^{\prime}}{aH^{2}}\right]\nonumber, \\
\gamma_{\text{D1}}\equiv\, & \left(\beta_{1}+\beta_{2}\right)^{2},
\end{align}
\begin{align}
G_{\text{N1}}\equiv\, & \frac{2\alpha_{\textrm{B}}\gamma_{\text{N1}}}{\alpha_{\textrm{M}}-\alpha_{\textrm{T}}}-2\left(6-6\alpha_{\textrm{B}}-\alpha_{\textrm{K}}\right)\left(\beta_{1}+\beta_{2}\right)\beta_{3}\\
 & +6\left(\beta_{1}+\beta_{2}\right)\beta_{2}\left[3\frac{\left(\rho_{\textrm{m}}+p_{\textrm{m}}\right)}{M_{*}^{2}H^{2}}+\left(2-\alpha_{\textrm{B}}\right)\frac{H^{\prime}}{aH^{2}}\right],\\
G_{\text{D1}}\equiv\, & M_{*}^{2}\left[2\beta_{1}+\left(2-\alpha_{\textrm{B}}\right)\beta_{2}\right]^{2},
\end{align}
where $G_{1}\equiv G_{\textrm{N}1}/G_{\textrm{D}1}$, $\gamma_{1}\equiv\gamma_{\textrm{N}1}/\gamma_{\textrm{D}2}$
and
\begin{align}\label{eq:betas}
\beta_{1}\equiv & -\frac{3\left(\rho_{\textrm{m}}+p_{\textrm{m}}\right)}{H^{2}M_{*}^{2}}+\frac{\alpha_{\textrm{B}}^{\prime}H-\left(2-\alpha_{\textrm{B}}\right)H^{\prime}}{aH^{2}},\\
\beta_{2}\equiv & \alpha_{\textrm{B}}\left(1+\alpha_{\textrm{T}}\right)+2\left(\alpha_{\textrm{M}}-\alpha_{\textrm{T}}\right),\\
\beta_{3}\equiv & \left(1+\alpha_{\textrm{T}}\right)\beta_{1}+\left(1+\alpha_{\textrm{M}}\right)\beta_{2}\,.
\end{align}
Here $'$ represents the derivative with respect to conformal time and ${\cal H}=aH$ and the normalisation of the densities is the one used in the CLASS code \cite{Blas:2011rf}.

\end{document}